\documentclass[longauth]{aa}

\usepackage{graphicx}
\usepackage{natbib}
\usepackage{txfonts,textcomp}
\usepackage{hyperref}
\hypersetup{
    colorlinks=true,
    linkcolor=blue,
    filecolor=magenta,      
    urlcolor=blue,
    citecolor=blue
}

\makeatletter
\renewcommand*\aa@pageof{, page \thepage{} of \pageref*{LastPage}}
\makeatother

\usepackage{euclid}
\usepackage{ISTL_macros}
\usepackage{bm}
\usepackage[capitalise,nameinlink]{cleveref}

\usepackage{glossaries}
\glsdisablehyper

\usepackage{orcidlink}
\newcommand{\orcid}[1]{\orcidlink{#1}}

\newacronym{lss}{LSS}{large-scale structure}
\newacronym{eft}{EFTofLSS}{effective field theory of large-scale structure}
\newacronym{rsd}{RSD}{Redshift space distortions}
\newacronym{gcsp}{GCsp}{spectroscopic galaxy clustering}
\newacronym{2pcf}{2PCF}{two-point correlation function}
\newacronym{mcmc}{MCMC}{Monte Carlo Markov chain}
\newacronym{gcph}{GCph}{photometric galaxy clustering}
\newacronym{wl}{WL}{weak lensing}
\newacronym{ggl}{GGL}{galaxy-galaxy lensing}

\defcitealias{EP-CLOE1}{Paper 1}
\defcitealias{Paper2}{Paper 2}
\defcitealias{EP-ClOE3}{Paper 3}

\begin{document}

\title{\Euclid preparation}
\subtitle{XCVII. Cosmology Likelihood for Observables in Euclid (CLOE). 4: Validation and performance}   											   

\author{Euclid Collaboration: M.~Martinelli\orcid{0000-0002-6943-7732}\thanks{\email{matteo.martinelli@inaf.it}}\inst{\ref{aff1},\ref{aff2}}
\and A.~Pezzotta\orcid{0000-0003-0726-2268}\inst{\ref{aff3},\ref{aff4}}
\and D.~Sciotti\orcid{0009-0008-4519-2620}\inst{\ref{aff1},\ref{aff2}}
\and L.~Blot\orcid{0000-0002-9622-7167}\inst{\ref{aff5},\ref{aff6}}
\and M.~Bonici\orcid{0000-0002-8430-126X}\inst{\ref{aff7},\ref{aff8}}
\and S.~Camera\orcid{0000-0003-3399-3574}\inst{\ref{aff9},\ref{aff10},\ref{aff11}}
\and G.~Ca\~nas-Herrera\orcid{0000-0003-2796-2149}\inst{\ref{aff12},\ref{aff13},\ref{aff14}}
\and V.~F.~Cardone\inst{\ref{aff1},\ref{aff2}}
\and P.~Carrilho\orcid{0000-0003-1339-0194}\inst{\ref{aff15}}
\and S.~Casas\orcid{0000-0002-4751-5138}\inst{\ref{aff16}}
\and S.~Davini\orcid{0000-0003-3269-1718}\inst{\ref{aff17}}
\and S.~Di~Domizio\orcid{0000-0003-2863-5895}\inst{\ref{aff18},\ref{aff17}}
\and S.~Farrens\orcid{0000-0002-9594-9387}\inst{\ref{aff19}}
\and L.~W.~K.~Goh\orcid{0000-0002-0104-8132}\inst{\ref{aff19}}
\and S.~Gouyou~Beauchamps\inst{\ref{aff20},\ref{aff21}}
\and S.~Ili\'c\orcid{0000-0003-4285-9086}\inst{\ref{aff22},\ref{aff23}}
\and S.~Joudaki\orcid{0000-0001-8820-673X}\inst{\ref{aff24},\ref{aff25}}
\and F.~Keil\orcid{0000-0002-8108-1679}\inst{\ref{aff23}}
\and A.~M.~C.~Le~Brun\orcid{0000-0002-0936-4594}\inst{\ref{aff26}}
\and C.~Moretti\orcid{0000-0003-3314-8936}\inst{\ref{aff27},\ref{aff28},\ref{aff29},\ref{aff30},\ref{aff31}}
\and V.~Pettorino\inst{\ref{aff12}}
\and A.~G.~S\'anchez\orcid{0000-0003-1198-831X}\inst{\ref{aff4}}
\and Z.~Sakr\orcid{0000-0002-4823-3757}\inst{\ref{aff32},\ref{aff23},\ref{aff33}}
\and K.~Tanidis\orcid{0000-0001-9843-5130}\inst{\ref{aff34}}
\and I.~Tutusaus\orcid{0000-0002-3199-0399}\inst{\ref{aff23}}
\and V.~Ajani\orcid{0000-0001-9442-2527}\inst{\ref{aff19},\ref{aff35},\ref{aff36}}
\and M.~Crocce\orcid{0000-0002-9745-6228}\inst{\ref{aff21},\ref{aff20}}
\and C.~Giocoli\orcid{0000-0002-9590-7961}\inst{\ref{aff37},\ref{aff38}}
\and L.~Legrand\orcid{0000-0003-0610-5252}\inst{\ref{aff39},\ref{aff40}}
\and M.~Lembo\orcid{0000-0002-5271-5070}\inst{\ref{aff41},\ref{aff42}}
\and G.~F.~Lesci\orcid{0000-0002-4607-2830}\inst{\ref{aff43},\ref{aff37}}
\and D.~Navarro-Giron\'es\orcid{0000-0003-0507-372X}\inst{\ref{aff14}}
\and A.~Nouri-Zonoz\orcid{0009-0006-6164-8670}\inst{\ref{aff44}}
\and S.~Pamuk\orcid{0009-0004-0852-8624}\inst{\ref{aff45}}
\and M.~Tsedrik\orcid{0000-0002-0020-5343}\inst{\ref{aff15},\ref{aff46}}
\and J.~Bel\inst{\ref{aff47}}
\and C.~Carbone\orcid{0000-0003-0125-3563}\inst{\ref{aff7}}
\and C.~A.~J.~Duncan\orcid{0009-0003-3573-0791}\inst{\ref{aff48}}
\and M.~Kilbinger\orcid{0000-0001-9513-7138}\inst{\ref{aff19}}
\and F.~Lacasa\orcid{0000-0002-7268-3440}\inst{\ref{aff49},\ref{aff50}}
\and M.~Lattanzi\orcid{0000-0003-1059-2532}\inst{\ref{aff42}}
\and D.~Sapone\orcid{0000-0001-7089-4503}\inst{\ref{aff51}}
\and E.~Sellentin\inst{\ref{aff52},\ref{aff14}}
\and P.~L.~Taylor\orcid{0000-0001-6999-4718}\inst{\ref{aff53},\ref{aff54}}
\and N.~Aghanim\orcid{0000-0002-6688-8992}\inst{\ref{aff50}}
\and B.~Altieri\orcid{0000-0003-3936-0284}\inst{\ref{aff55}}
\and L.~Amendola\orcid{0000-0002-0835-233X}\inst{\ref{aff32}}
\and S.~Andreon\orcid{0000-0002-2041-8784}\inst{\ref{aff3}}
\and N.~Auricchio\orcid{0000-0003-4444-8651}\inst{\ref{aff37}}
\and C.~Baccigalupi\orcid{0000-0002-8211-1630}\inst{\ref{aff30},\ref{aff29},\ref{aff31},\ref{aff27}}
\and M.~Baldi\orcid{0000-0003-4145-1943}\inst{\ref{aff56},\ref{aff37},\ref{aff38}}
\and A.~Balestra\orcid{0000-0002-6967-261X}\inst{\ref{aff57}}
\and S.~Bardelli\orcid{0000-0002-8900-0298}\inst{\ref{aff37}}
\and P.~Battaglia\orcid{0000-0002-7337-5909}\inst{\ref{aff37}}
\and R.~Bender\orcid{0000-0001-7179-0626}\inst{\ref{aff4},\ref{aff58}}
\and A.~Biviano\orcid{0000-0002-0857-0732}\inst{\ref{aff29},\ref{aff30}}
\and A.~Bonchi\orcid{0000-0002-2667-5482}\inst{\ref{aff59}}
\and D.~Bonino\orcid{0000-0002-3336-9977}\inst{\ref{aff11}}
\and E.~Branchini\orcid{0000-0002-0808-6908}\inst{\ref{aff18},\ref{aff17},\ref{aff3}}
\and M.~Brescia\orcid{0000-0001-9506-5680}\inst{\ref{aff60},\ref{aff61}}
\and J.~Brinchmann\orcid{0000-0003-4359-8797}\inst{\ref{aff62},\ref{aff63}}
\and V.~Capobianco\orcid{0000-0002-3309-7692}\inst{\ref{aff11}}
\and J.~Carretero\orcid{0000-0002-3130-0204}\inst{\ref{aff24},\ref{aff64}}
\and M.~Castellano\orcid{0000-0001-9875-8263}\inst{\ref{aff1}}
\and G.~Castignani\orcid{0000-0001-6831-0687}\inst{\ref{aff37}}
\and S.~Cavuoti\orcid{0000-0002-3787-4196}\inst{\ref{aff61},\ref{aff65}}
\and K.~C.~Chambers\orcid{0000-0001-6965-7789}\inst{\ref{aff66}}
\and A.~Cimatti\inst{\ref{aff67}}
\and C.~Colodro-Conde\inst{\ref{aff68}}
\and G.~Congedo\orcid{0000-0003-2508-0046}\inst{\ref{aff15}}
\and C.~J.~Conselice\orcid{0000-0003-1949-7638}\inst{\ref{aff48}}
\and L.~Conversi\orcid{0000-0002-6710-8476}\inst{\ref{aff69},\ref{aff55}}
\and Y.~Copin\orcid{0000-0002-5317-7518}\inst{\ref{aff70}}
\and H.~M.~Courtois\orcid{0000-0003-0509-1776}\inst{\ref{aff71}}
\and A.~Da~Silva\orcid{0000-0002-6385-1609}\inst{\ref{aff72},\ref{aff73}}
\and H.~Degaudenzi\orcid{0000-0002-5887-6799}\inst{\ref{aff74}}
\and S.~de~la~Torre\inst{\ref{aff75}}
\and G.~De~Lucia\orcid{0000-0002-6220-9104}\inst{\ref{aff29}}
\and A.~M.~Di~Giorgio\orcid{0000-0002-4767-2360}\inst{\ref{aff76}}
\and H.~Dole\orcid{0000-0002-9767-3839}\inst{\ref{aff50}}
\and F.~Dubath\orcid{0000-0002-6533-2810}\inst{\ref{aff74}}
\and X.~Dupac\inst{\ref{aff55}}
\and S.~Dusini\orcid{0000-0002-1128-0664}\inst{\ref{aff77}}
\and A.~Ealet\orcid{0000-0003-3070-014X}\inst{\ref{aff70}}
\and S.~Escoffier\orcid{0000-0002-2847-7498}\inst{\ref{aff78}}
\and M.~Farina\orcid{0000-0002-3089-7846}\inst{\ref{aff76}}
\and R.~Farinelli\inst{\ref{aff37}}
\and F.~Faustini\orcid{0000-0001-6274-5145}\inst{\ref{aff59},\ref{aff1}}
\and S.~Ferriol\inst{\ref{aff70}}
\and F.~Finelli\orcid{0000-0002-6694-3269}\inst{\ref{aff37},\ref{aff79}}
\and P.~Fosalba\orcid{0000-0002-1510-5214}\inst{\ref{aff20},\ref{aff21}}
\and S.~Fotopoulou\orcid{0000-0002-9686-254X}\inst{\ref{aff80}}
\and N.~Fourmanoit\orcid{0009-0005-6816-6925}\inst{\ref{aff78}}
\and M.~Frailis\orcid{0000-0002-7400-2135}\inst{\ref{aff29}}
\and E.~Franceschi\orcid{0000-0002-0585-6591}\inst{\ref{aff37}}
\and M.~Fumana\orcid{0000-0001-6787-5950}\inst{\ref{aff7}}
\and S.~Galeotta\orcid{0000-0002-3748-5115}\inst{\ref{aff29}}
\and K.~George\orcid{0000-0002-1734-8455}\inst{\ref{aff58}}
\and W.~Gillard\orcid{0000-0003-4744-9748}\inst{\ref{aff78}}
\and B.~Gillis\orcid{0000-0002-4478-1270}\inst{\ref{aff15}}
\and J.~Gracia-Carpio\inst{\ref{aff4}}
\and B.~R.~Granett\orcid{0000-0003-2694-9284}\inst{\ref{aff3}}
\and A.~Grazian\orcid{0000-0002-5688-0663}\inst{\ref{aff57}}
\and F.~Grupp\inst{\ref{aff4},\ref{aff58}}
\and L.~Guzzo\orcid{0000-0001-8264-5192}\inst{\ref{aff81},\ref{aff3},\ref{aff82}}
\and S.~V.~H.~Haugan\orcid{0000-0001-9648-7260}\inst{\ref{aff83}}
\and H.~Hoekstra\orcid{0000-0002-0641-3231}\inst{\ref{aff14}}
\and W.~Holmes\inst{\ref{aff84}}
\and F.~Hormuth\inst{\ref{aff85}}
\and A.~Hornstrup\orcid{0000-0002-3363-0936}\inst{\ref{aff86},\ref{aff87}}
\and P.~Hudelot\inst{\ref{aff88}}
\and K.~Jahnke\orcid{0000-0003-3804-2137}\inst{\ref{aff89}}
\and M.~Jhabvala\inst{\ref{aff90}}
\and B.~Joachimi\orcid{0000-0001-7494-1303}\inst{\ref{aff91}}
\and E.~Keih\"anen\orcid{0000-0003-1804-7715}\inst{\ref{aff92}}
\and S.~Kermiche\orcid{0000-0002-0302-5735}\inst{\ref{aff78}}
\and A.~Kiessling\orcid{0000-0002-2590-1273}\inst{\ref{aff84}}
\and B.~Kubik\orcid{0009-0006-5823-4880}\inst{\ref{aff70}}
\and K.~Kuijken\orcid{0000-0002-3827-0175}\inst{\ref{aff14}}
\and M.~K\"ummel\orcid{0000-0003-2791-2117}\inst{\ref{aff58}}
\and M.~Kunz\orcid{0000-0002-3052-7394}\inst{\ref{aff44}}
\and H.~Kurki-Suonio\orcid{0000-0002-4618-3063}\inst{\ref{aff93},\ref{aff94}}
\and P.~Liebing\inst{\ref{aff95}}
\and S.~Ligori\orcid{0000-0003-4172-4606}\inst{\ref{aff11}}
\and P.~B.~Lilje\orcid{0000-0003-4324-7794}\inst{\ref{aff83}}
\and V.~Lindholm\orcid{0000-0003-2317-5471}\inst{\ref{aff93},\ref{aff94}}
\and I.~Lloro\orcid{0000-0001-5966-1434}\inst{\ref{aff96}}
\and G.~Mainetti\orcid{0000-0003-2384-2377}\inst{\ref{aff97}}
\and D.~Maino\inst{\ref{aff81},\ref{aff7},\ref{aff82}}
\and E.~Maiorano\orcid{0000-0003-2593-4355}\inst{\ref{aff37}}
\and O.~Mansutti\orcid{0000-0001-5758-4658}\inst{\ref{aff29}}
\and S.~Marcin\inst{\ref{aff98}}
\and O.~Marggraf\orcid{0000-0001-7242-3852}\inst{\ref{aff99}}
\and K.~Markovic\orcid{0000-0001-6764-073X}\inst{\ref{aff84}}
\and N.~Martinet\orcid{0000-0003-2786-7790}\inst{\ref{aff75}}
\and F.~Marulli\orcid{0000-0002-8850-0303}\inst{\ref{aff43},\ref{aff37},\ref{aff38}}
\and R.~Massey\orcid{0000-0002-6085-3780}\inst{\ref{aff100}}
\and S.~Maurogordato\inst{\ref{aff101}}
\and E.~Medinaceli\orcid{0000-0002-4040-7783}\inst{\ref{aff37}}
\and S.~Mei\orcid{0000-0002-2849-559X}\inst{\ref{aff102},\ref{aff103}}
\and Y.~Mellier\inst{\ref{aff104},\ref{aff88}}
\and M.~Meneghetti\orcid{0000-0003-1225-7084}\inst{\ref{aff37},\ref{aff38}}
\and E.~Merlin\orcid{0000-0001-6870-8900}\inst{\ref{aff1}}
\and G.~Meylan\inst{\ref{aff105}}
\and A.~Mora\orcid{0000-0002-1922-8529}\inst{\ref{aff106}}
\and M.~Moresco\orcid{0000-0002-7616-7136}\inst{\ref{aff43},\ref{aff37}}
\and L.~Moscardini\orcid{0000-0002-3473-6716}\inst{\ref{aff43},\ref{aff37},\ref{aff38}}
\and C.~Neissner\orcid{0000-0001-8524-4968}\inst{\ref{aff107},\ref{aff64}}
\and S.-M.~Niemi\inst{\ref{aff12}}
\and J.~W.~Nightingale\orcid{0000-0002-8987-7401}\inst{\ref{aff108}}
\and C.~Padilla\orcid{0000-0001-7951-0166}\inst{\ref{aff107}}
\and S.~Paltani\orcid{0000-0002-8108-9179}\inst{\ref{aff74}}
\and F.~Pasian\orcid{0000-0002-4869-3227}\inst{\ref{aff29}}
\and K.~Pedersen\inst{\ref{aff109}}
\and W.~J.~Percival\orcid{0000-0002-0644-5727}\inst{\ref{aff8},\ref{aff110},\ref{aff111}}
\and S.~Pires\orcid{0000-0002-0249-2104}\inst{\ref{aff19}}
\and G.~Polenta\orcid{0000-0003-4067-9196}\inst{\ref{aff59}}
\and M.~Poncet\inst{\ref{aff112}}
\and L.~A.~Popa\inst{\ref{aff113}}
\and F.~Raison\orcid{0000-0002-7819-6918}\inst{\ref{aff4}}
\and R.~Rebolo\inst{\ref{aff68},\ref{aff114},\ref{aff115}}
\and A.~Renzi\orcid{0000-0001-9856-1970}\inst{\ref{aff116},\ref{aff77}}
\and J.~Rhodes\orcid{0000-0002-4485-8549}\inst{\ref{aff84}}
\and G.~Riccio\inst{\ref{aff61}}
\and E.~Romelli\orcid{0000-0003-3069-9222}\inst{\ref{aff29}}
\and M.~Roncarelli\orcid{0000-0001-9587-7822}\inst{\ref{aff37}}
\and R.~Saglia\orcid{0000-0003-0378-7032}\inst{\ref{aff58},\ref{aff4}}
\and B.~Sartoris\orcid{0000-0003-1337-5269}\inst{\ref{aff58},\ref{aff29}}
\and J.~A.~Schewtschenko\orcid{0000-0002-4913-6393}\inst{\ref{aff15}}
\and P.~Schneider\orcid{0000-0001-8561-2679}\inst{\ref{aff99}}
\and T.~Schrabback\orcid{0000-0002-6987-7834}\inst{\ref{aff117}}
\and A.~Secroun\orcid{0000-0003-0505-3710}\inst{\ref{aff78}}
\and E.~Sefusatti\orcid{0000-0003-0473-1567}\inst{\ref{aff29},\ref{aff30},\ref{aff31}}
\and G.~Seidel\orcid{0000-0003-2907-353X}\inst{\ref{aff89}}
\and M.~Seiffert\orcid{0000-0002-7536-9393}\inst{\ref{aff84}}
\and S.~Serrano\orcid{0000-0002-0211-2861}\inst{\ref{aff20},\ref{aff118},\ref{aff21}}
\and P.~Simon\inst{\ref{aff99}}
\and C.~Sirignano\orcid{0000-0002-0995-7146}\inst{\ref{aff116},\ref{aff77}}
\and G.~Sirri\orcid{0000-0003-2626-2853}\inst{\ref{aff38}}
\and A.~Spurio~Mancini\orcid{0000-0001-5698-0990}\inst{\ref{aff119}}
\and L.~Stanco\orcid{0000-0002-9706-5104}\inst{\ref{aff77}}
\and J.~Steinwagner\orcid{0000-0001-7443-1047}\inst{\ref{aff4}}
\and P.~Tallada-Cresp\'{i}\orcid{0000-0002-1336-8328}\inst{\ref{aff24},\ref{aff64}}
\and D.~Tavagnacco\orcid{0000-0001-7475-9894}\inst{\ref{aff29}}
\and A.~N.~Taylor\inst{\ref{aff15}}
\and I.~Tereno\inst{\ref{aff72},\ref{aff120}}
\and S.~Toft\orcid{0000-0003-3631-7176}\inst{\ref{aff121},\ref{aff122}}
\and R.~Toledo-Moreo\orcid{0000-0002-2997-4859}\inst{\ref{aff123}}
\and F.~Torradeflot\orcid{0000-0003-1160-1517}\inst{\ref{aff64},\ref{aff24}}
\and L.~Valenziano\orcid{0000-0002-1170-0104}\inst{\ref{aff37},\ref{aff79}}
\and J.~Valiviita\orcid{0000-0001-6225-3693}\inst{\ref{aff93},\ref{aff94}}
\and T.~Vassallo\orcid{0000-0001-6512-6358}\inst{\ref{aff58},\ref{aff29}}
\and G.~Verdoes~Kleijn\orcid{0000-0001-5803-2580}\inst{\ref{aff124}}
\and A.~Veropalumbo\orcid{0000-0003-2387-1194}\inst{\ref{aff3},\ref{aff17},\ref{aff18}}
\and Y.~Wang\orcid{0000-0002-4749-2984}\inst{\ref{aff125}}
\and J.~Weller\orcid{0000-0002-8282-2010}\inst{\ref{aff58},\ref{aff4}}
\and G.~Zamorani\orcid{0000-0002-2318-301X}\inst{\ref{aff37}}
\and F.~M.~Zerbi\inst{\ref{aff3}}
\and E.~Zucca\orcid{0000-0002-5845-8132}\inst{\ref{aff37}}
\and V.~Allevato\orcid{0000-0001-7232-5152}\inst{\ref{aff61}}
\and M.~Ballardini\orcid{0000-0003-4481-3559}\inst{\ref{aff41},\ref{aff42},\ref{aff37}}
\and M.~Bolzonella\orcid{0000-0003-3278-4607}\inst{\ref{aff37}}
\and E.~Bozzo\orcid{0000-0002-8201-1525}\inst{\ref{aff74}}
\and C.~Burigana\orcid{0000-0002-3005-5796}\inst{\ref{aff126},\ref{aff79}}
\and R.~Cabanac\orcid{0000-0001-6679-2600}\inst{\ref{aff23}}
\and M.~Calabrese\orcid{0000-0002-2637-2422}\inst{\ref{aff127},\ref{aff7}}
\and D.~Di~Ferdinando\inst{\ref{aff38}}
\and J.~A.~Escartin~Vigo\inst{\ref{aff4}}
\and L.~Gabarra\orcid{0000-0002-8486-8856}\inst{\ref{aff34}}
\and J.~Mart\'{i}n-Fleitas\orcid{0000-0002-8594-569X}\inst{\ref{aff106}}
\and S.~Matthew\orcid{0000-0001-8448-1697}\inst{\ref{aff15}}
\and N.~Mauri\orcid{0000-0001-8196-1548}\inst{\ref{aff67},\ref{aff38}}
\and R.~B.~Metcalf\orcid{0000-0003-3167-2574}\inst{\ref{aff43},\ref{aff37}}
\and M.~P\"ontinen\orcid{0000-0001-5442-2530}\inst{\ref{aff93}}
\and C.~Porciani\orcid{0000-0002-7797-2508}\inst{\ref{aff99}}
\and I.~Risso\orcid{0000-0003-2525-7761}\inst{\ref{aff128}}
\and V.~Scottez\inst{\ref{aff104},\ref{aff129}}
\and M.~Sereno\orcid{0000-0003-0302-0325}\inst{\ref{aff37},\ref{aff38}}
\and M.~Tenti\orcid{0000-0002-4254-5901}\inst{\ref{aff38}}
\and M.~Viel\orcid{0000-0002-2642-5707}\inst{\ref{aff30},\ref{aff29},\ref{aff27},\ref{aff31},\ref{aff28}}
\and M.~Wiesmann\orcid{0009-0000-8199-5860}\inst{\ref{aff83}}
\and Y.~Akrami\orcid{0000-0002-2407-7956}\inst{\ref{aff130},\ref{aff131}}
\and S.~Alvi\orcid{0000-0001-5779-8568}\inst{\ref{aff41}}
\and I.~T.~Andika\orcid{0000-0001-6102-9526}\inst{\ref{aff132},\ref{aff133}}
\and R.~E.~Angulo\orcid{0000-0003-2953-3970}\inst{\ref{aff134},\ref{aff135}}
\and S.~Anselmi\orcid{0000-0002-3579-9583}\inst{\ref{aff77},\ref{aff116},\ref{aff6}}
\and M.~Archidiacono\orcid{0000-0003-4952-9012}\inst{\ref{aff81},\ref{aff82}}
\and F.~Atrio-Barandela\orcid{0000-0002-2130-2513}\inst{\ref{aff136}}
\and A.~Balaguera-Antolinez\orcid{0000-0001-5028-3035}\inst{\ref{aff68}}
\and M.~Bethermin\orcid{0000-0002-3915-2015}\inst{\ref{aff137}}
\and A.~Blanchard\orcid{0000-0001-8555-9003}\inst{\ref{aff23}}
\and H.~B\"ohringer\orcid{0000-0001-8241-4204}\inst{\ref{aff4},\ref{aff138},\ref{aff139}}
\and S.~Borgani\orcid{0000-0001-6151-6439}\inst{\ref{aff140},\ref{aff30},\ref{aff29},\ref{aff31},\ref{aff28}}
\and M.~L.~Brown\orcid{0000-0002-0370-8077}\inst{\ref{aff48}}
\and S.~Bruton\orcid{0000-0002-6503-5218}\inst{\ref{aff141}}
\and A.~Calabro\orcid{0000-0003-2536-1614}\inst{\ref{aff1}}
\and B.~Camacho~Quevedo\orcid{0000-0002-8789-4232}\inst{\ref{aff20},\ref{aff21}}
\and A.~Cappi\inst{\ref{aff37},\ref{aff101}}
\and F.~Caro\inst{\ref{aff1}}
\and C.~S.~Carvalho\inst{\ref{aff120}}
\and T.~Castro\orcid{0000-0002-6292-3228}\inst{\ref{aff29},\ref{aff31},\ref{aff30},\ref{aff28}}
\and F.~Cogato\orcid{0000-0003-4632-6113}\inst{\ref{aff43},\ref{aff37}}
\and S.~Conseil\orcid{0000-0002-3657-4191}\inst{\ref{aff70}}
\and A.~R.~Cooray\orcid{0000-0002-3892-0190}\inst{\ref{aff142}}
\and O.~Cucciati\orcid{0000-0002-9336-7551}\inst{\ref{aff37}}
\and F.~De~Paolis\orcid{0000-0001-6460-7563}\inst{\ref{aff143},\ref{aff144},\ref{aff145}}
\and G.~Desprez\orcid{0000-0001-8325-1742}\inst{\ref{aff124}}
\and A.~D\'iaz-S\'anchez\orcid{0000-0003-0748-4768}\inst{\ref{aff146}}
\and J.~J.~Diaz\inst{\ref{aff68}}
\and J.~M.~Diego\orcid{0000-0001-9065-3926}\inst{\ref{aff45}}
\and P.~Dimauro\orcid{0000-0001-7399-2854}\inst{\ref{aff1},\ref{aff147}}
\and A.~Enia\orcid{0000-0002-0200-2857}\inst{\ref{aff56},\ref{aff37}}
\and Y.~Fang\inst{\ref{aff58}}
\and A.~G.~Ferrari\orcid{0009-0005-5266-4110}\inst{\ref{aff38}}
\and P.~G.~Ferreira\orcid{0000-0002-3021-2851}\inst{\ref{aff34}}
\and A.~Finoguenov\orcid{0000-0002-4606-5403}\inst{\ref{aff93}}
\and A.~Fontana\orcid{0000-0003-3820-2823}\inst{\ref{aff1}}
\and A.~Franco\orcid{0000-0002-4761-366X}\inst{\ref{aff144},\ref{aff143},\ref{aff145}}
\and K.~Ganga\orcid{0000-0001-8159-8208}\inst{\ref{aff102}}
\and J.~Garc\'ia-Bellido\orcid{0000-0002-9370-8360}\inst{\ref{aff130}}
\and T.~Gasparetto\orcid{0000-0002-7913-4866}\inst{\ref{aff29}}
\and V.~Gautard\inst{\ref{aff148}}
\and R.~Gavazzi\orcid{0000-0002-5540-6935}\inst{\ref{aff75},\ref{aff88}}
\and E.~Gaztanaga\orcid{0000-0001-9632-0815}\inst{\ref{aff21},\ref{aff20},\ref{aff25}}
\and F.~Giacomini\orcid{0000-0002-3129-2814}\inst{\ref{aff38}}
\and F.~Gianotti\orcid{0000-0003-4666-119X}\inst{\ref{aff37}}
\and G.~Gozaliasl\orcid{0000-0002-0236-919X}\inst{\ref{aff149},\ref{aff93}}
\and A.~Gruppuso\orcid{0000-0001-9272-5292}\inst{\ref{aff37},\ref{aff38}}
\and M.~Guidi\orcid{0000-0001-9408-1101}\inst{\ref{aff56},\ref{aff37}}
\and C.~M.~Gutierrez\orcid{0000-0001-7854-783X}\inst{\ref{aff150}}
\and A.~Hall\orcid{0000-0002-3139-8651}\inst{\ref{aff15}}
\and S.~Hemmati\orcid{0000-0003-2226-5395}\inst{\ref{aff151}}
\and H.~Hildebrandt\orcid{0000-0002-9814-3338}\inst{\ref{aff152}}
\and J.~Hjorth\orcid{0000-0002-4571-2306}\inst{\ref{aff109}}
\and J.~J.~E.~Kajava\orcid{0000-0002-3010-8333}\inst{\ref{aff153},\ref{aff154}}
\and Y.~Kang\orcid{0009-0000-8588-7250}\inst{\ref{aff74}}
\and V.~Kansal\orcid{0000-0002-4008-6078}\inst{\ref{aff155},\ref{aff156}}
\and D.~Karagiannis\orcid{0000-0002-4927-0816}\inst{\ref{aff41},\ref{aff157}}
\and K.~Kiiveri\inst{\ref{aff92}}
\and C.~C.~Kirkpatrick\inst{\ref{aff92}}
\and S.~Kruk\orcid{0000-0001-8010-8879}\inst{\ref{aff55}}
\and J.~Le~Graet\orcid{0000-0001-6523-7971}\inst{\ref{aff78}}
\and F.~Lepori\orcid{0009-0000-5061-7138}\inst{\ref{aff158}}
\and G.~Leroy\orcid{0009-0004-2523-4425}\inst{\ref{aff159},\ref{aff100}}
\and J.~Lesgourgues\orcid{0000-0001-7627-353X}\inst{\ref{aff16}}
\and L.~Leuzzi\orcid{0009-0006-4479-7017}\inst{\ref{aff43},\ref{aff37}}
\and T.~I.~Liaudat\orcid{0000-0002-9104-314X}\inst{\ref{aff160}}
\and S.~J.~Liu\orcid{0000-0001-7680-2139}\inst{\ref{aff76}}
\and A.~Loureiro\orcid{0000-0002-4371-0876}\inst{\ref{aff161},\ref{aff162}}
\and J.~Macias-Perez\orcid{0000-0002-5385-2763}\inst{\ref{aff163}}
\and G.~Maggio\orcid{0000-0003-4020-4836}\inst{\ref{aff29}}
\and M.~Magliocchetti\orcid{0000-0001-9158-4838}\inst{\ref{aff76}}
\and F.~Mannucci\orcid{0000-0002-4803-2381}\inst{\ref{aff164}}
\and R.~Maoli\orcid{0000-0002-6065-3025}\inst{\ref{aff165},\ref{aff1}}
\and C.~J.~A.~P.~Martins\orcid{0000-0002-4886-9261}\inst{\ref{aff166},\ref{aff62}}
\and L.~Maurin\orcid{0000-0002-8406-0857}\inst{\ref{aff50}}
\and M.~Migliaccio\inst{\ref{aff167},\ref{aff168}}
\and M.~Miluzio\inst{\ref{aff55},\ref{aff169}}
\and P.~Monaco\orcid{0000-0003-2083-7564}\inst{\ref{aff140},\ref{aff29},\ref{aff31},\ref{aff30}}
\and G.~Morgante\inst{\ref{aff37}}
\and S.~Nadathur\orcid{0000-0001-9070-3102}\inst{\ref{aff25}}
\and K.~Naidoo\orcid{0000-0002-9182-1802}\inst{\ref{aff25}}
\and P.~Natoli\orcid{0000-0003-0126-9100}\inst{\ref{aff41},\ref{aff42}}
\and A.~Navarro-Alsina\orcid{0000-0002-3173-2592}\inst{\ref{aff99}}
\and S.~Nesseris\orcid{0000-0002-0567-0324}\inst{\ref{aff130}}
\and L.~Pagano\orcid{0000-0003-1820-5998}\inst{\ref{aff41},\ref{aff42}}
\and F.~Passalacqua\orcid{0000-0002-8606-4093}\inst{\ref{aff116},\ref{aff77}}
\and K.~Paterson\orcid{0000-0001-8340-3486}\inst{\ref{aff89}}
\and L.~Patrizii\inst{\ref{aff38}}
\and A.~Pisani\orcid{0000-0002-6146-4437}\inst{\ref{aff78},\ref{aff170}}
\and D.~Potter\orcid{0000-0002-0757-5195}\inst{\ref{aff158}}
\and S.~Quai\orcid{0000-0002-0449-8163}\inst{\ref{aff43},\ref{aff37}}
\and M.~Radovich\orcid{0000-0002-3585-866X}\inst{\ref{aff57}}
\and P.-F.~Rocci\inst{\ref{aff50}}
\and S.~Sacquegna\orcid{0000-0002-8433-6630}\inst{\ref{aff143},\ref{aff144},\ref{aff145}}
\and M.~Sahl\'en\orcid{0000-0003-0973-4804}\inst{\ref{aff171}}
\and D.~B.~Sanders\orcid{0000-0002-1233-9998}\inst{\ref{aff66}}
\and E.~Sarpa\orcid{0000-0002-1256-655X}\inst{\ref{aff27},\ref{aff28},\ref{aff31}}
\and A.~Schneider\orcid{0000-0001-7055-8104}\inst{\ref{aff158}}
\and A.~Shulevski\orcid{0000-0002-1827-0469}\inst{\ref{aff172},\ref{aff124},\ref{aff173},\ref{aff174}}
\and A.~Silvestri\orcid{0000-0001-6904-5061}\inst{\ref{aff13}}
\and L.~C.~Smith\orcid{0000-0002-3259-2771}\inst{\ref{aff175}}
\and J.~Stadel\orcid{0000-0001-7565-8622}\inst{\ref{aff158}}
\and C.~Tao\orcid{0000-0001-7961-8177}\inst{\ref{aff78}}
\and G.~Testera\inst{\ref{aff17}}
\and R.~Teyssier\orcid{0000-0001-7689-0933}\inst{\ref{aff170}}
\and S.~Tosi\orcid{0000-0002-7275-9193}\inst{\ref{aff18},\ref{aff128}}
\and A.~Troja\orcid{0000-0003-0239-4595}\inst{\ref{aff116},\ref{aff77}}
\and M.~Tucci\inst{\ref{aff74}}
\and C.~Valieri\inst{\ref{aff38}}
\and A.~Venhola\orcid{0000-0001-6071-4564}\inst{\ref{aff176}}
\and D.~Vergani\orcid{0000-0003-0898-2216}\inst{\ref{aff37}}
\and F.~Vernizzi\orcid{0000-0003-3426-2802}\inst{\ref{aff177}}
\and G.~Verza\orcid{0000-0002-1886-8348}\inst{\ref{aff178}}
\and P.~Vielzeuf\orcid{0000-0003-2035-9339}\inst{\ref{aff78}}
\and N.~A.~Walton\orcid{0000-0003-3983-8778}\inst{\ref{aff175}}}

\institute{INAF-Osservatorio Astronomico di Roma, Via Frascati 33, 00078 Monteporzio Catone, Italy\label{aff1}
\and
INFN-Sezione di Roma, Piazzale Aldo Moro, 2 - c/o Dipartimento di Fisica, Edificio G. Marconi, 00185 Roma, Italy\label{aff2}
\and
INAF-Osservatorio Astronomico di Brera, Via Brera 28, 20122 Milano, Italy\label{aff3}
\and
Max Planck Institute for Extraterrestrial Physics, Giessenbachstr. 1, 85748 Garching, Germany\label{aff4}
\and
Center for Data-Driven Discovery, Kavli IPMU (WPI), UTIAS, The University of Tokyo, Kashiwa, Chiba 277-8583, Japan\label{aff5}
\and
Laboratoire Univers et Th\'eorie, Observatoire de Paris, Universit\'e PSL, Universit\'e Paris Cit\'e, CNRS, 92190 Meudon, France\label{aff6}
\and
INAF-IASF Milano, Via Alfonso Corti 12, 20133 Milano, Italy\label{aff7}
\and
Waterloo Centre for Astrophysics, University of Waterloo, Waterloo, Ontario N2L 3G1, Canada\label{aff8}
\and
Dipartimento di Fisica, Universit\`a degli Studi di Torino, Via P. Giuria 1, 10125 Torino, Italy\label{aff9}
\and
INFN-Sezione di Torino, Via P. Giuria 1, 10125 Torino, Italy\label{aff10}
\and
INAF-Osservatorio Astrofisico di Torino, Via Osservatorio 20, 10025 Pino Torinese (TO), Italy\label{aff11}
\and
European Space Agency/ESTEC, Keplerlaan 1, 2201 AZ Noordwijk, The Netherlands\label{aff12}
\and
Institute Lorentz, Leiden University, Niels Bohrweg 2, 2333 CA Leiden, The Netherlands\label{aff13}
\and
Leiden Observatory, Leiden University, Einsteinweg 55, 2333 CC Leiden, The Netherlands\label{aff14}
\and
Institute for Astronomy, University of Edinburgh, Royal Observatory, Blackford Hill, Edinburgh EH9 3HJ, UK\label{aff15}
\and
Institute for Theoretical Particle Physics and Cosmology (TTK), RWTH Aachen University, 52056 Aachen, Germany\label{aff16}
\and
INFN-Sezione di Genova, Via Dodecaneso 33, 16146, Genova, Italy\label{aff17}
\and
Dipartimento di Fisica, Universit\`a di Genova, Via Dodecaneso 33, 16146, Genova, Italy\label{aff18}
\and
Universit\'e Paris-Saclay, Universit\'e Paris Cit\'e, CEA, CNRS, AIM, 91191, Gif-sur-Yvette, France\label{aff19}
\and
Institut d'Estudis Espacials de Catalunya (IEEC),  Edifici RDIT, Campus UPC, 08860 Castelldefels, Barcelona, Spain\label{aff20}
\and
Institute of Space Sciences (ICE, CSIC), Campus UAB, Carrer de Can Magrans, s/n, 08193 Barcelona, Spain\label{aff21}
\and
Universit\'e Paris-Saclay, CNRS/IN2P3, IJCLab, 91405 Orsay, France\label{aff22}
\and
Institut de Recherche en Astrophysique et Plan\'etologie (IRAP), Universit\'e de Toulouse, CNRS, UPS, CNES, 14 Av. Edouard Belin, 31400 Toulouse, France\label{aff23}
\and
Centro de Investigaciones Energ\'eticas, Medioambientales y Tecnol\'ogicas (CIEMAT), Avenida Complutense 40, 28040 Madrid, Spain\label{aff24}
\and
Institute of Cosmology and Gravitation, University of Portsmouth, Portsmouth PO1 3FX, UK\label{aff25}
\and
Laboratoire d'etude de l'Univers et des phenomenes eXtremes, Observatoire de Paris, Universit\'e PSL, Sorbonne Universit\'e, CNRS, 92190 Meudon, France\label{aff26}
\and
SISSA, International School for Advanced Studies, Via Bonomea 265, 34136 Trieste TS, Italy\label{aff27}
\and
ICSC - Centro Nazionale di Ricerca in High Performance Computing, Big Data e Quantum Computing, Via Magnanelli 2, Bologna, Italy\label{aff28}
\and
INAF-Osservatorio Astronomico di Trieste, Via G. B. Tiepolo 11, 34143 Trieste, Italy\label{aff29}
\and
IFPU, Institute for Fundamental Physics of the Universe, via Beirut 2, 34151 Trieste, Italy\label{aff30}
\and
INFN, Sezione di Trieste, Via Valerio 2, 34127 Trieste TS, Italy\label{aff31}
\and
Institut f\"ur Theoretische Physik, University of Heidelberg, Philosophenweg 16, 69120 Heidelberg, Germany\label{aff32}
\and
Universit\'e St Joseph; Faculty of Sciences, Beirut, Lebanon\label{aff33}
\and
Department of Physics, Oxford University, Keble Road, Oxford OX1 3RH, UK\label{aff34}
\and
Institute for Particle Physics and Astrophysics, Dept. of Physics, ETH Zurich, Wolfgang-Pauli-Strasse 27, 8093 Zurich, Switzerland\label{aff35}
\and
LINKS Foundation, Via Pier Carlo Boggio, 61 10138 Torino, Italy\label{aff36}
\and
INAF-Osservatorio di Astrofisica e Scienza dello Spazio di Bologna, Via Piero Gobetti 93/3, 40129 Bologna, Italy\label{aff37}
\and
INFN-Sezione di Bologna, Viale Berti Pichat 6/2, 40127 Bologna, Italy\label{aff38}
\and
DAMTP, Centre for Mathematical Sciences, Wilberforce Road, Cambridge CB3 0WA, UK\label{aff39}
\and
Kavli Institute for Cosmology Cambridge, Madingley Road, Cambridge, CB3 0HA, UK\label{aff40}
\and
Dipartimento di Fisica e Scienze della Terra, Universit\`a degli Studi di Ferrara, Via Giuseppe Saragat 1, 44122 Ferrara, Italy\label{aff41}
\and
Istituto Nazionale di Fisica Nucleare, Sezione di Ferrara, Via Giuseppe Saragat 1, 44122 Ferrara, Italy\label{aff42}
\and
Dipartimento di Fisica e Astronomia "Augusto Righi" - Alma Mater Studiorum Universit\`a di Bologna, via Piero Gobetti 93/2, 40129 Bologna, Italy\label{aff43}
\and
Universit\'e de Gen\`eve, D\'epartement de Physique Th\'eorique and Centre for Astroparticle Physics, 24 quai Ernest-Ansermet, CH-1211 Gen\`eve 4, Switzerland\label{aff44}
\and
Instituto de F\'isica de Cantabria, Edificio Juan Jord\'a, Avenida de los Castros, 39005 Santander, Spain\label{aff45}
\and
Higgs Centre for Theoretical Physics, School of Physics and Astronomy, The University of Edinburgh, Edinburgh EH9 3FD, UK\label{aff46}
\and
Aix-Marseille Universit\'e, Universit\'e de Toulon, CNRS, CPT, Marseille, France\label{aff47}
\and
Jodrell Bank Centre for Astrophysics, Department of Physics and Astronomy, University of Manchester, Oxford Road, Manchester M13 9PL, UK\label{aff48}
\and
Universit\'e Libre de Bruxelles (ULB), Service de Physique Th\'eorique CP225, Boulevard du Triophe, 1050 Bruxelles, Belgium\label{aff49}
\and
Universit\'e Paris-Saclay, CNRS, Institut d'astrophysique spatiale, 91405, Orsay, France\label{aff50}
\and
Departamento de F\'isica, FCFM, Universidad de Chile, Blanco Encalada 2008, Santiago, Chile\label{aff51}
\and
Mathematical Institute, University of Leiden, Einsteinweg 55, 2333 CA Leiden, The Netherlands\label{aff52}
\and
Center for Cosmology and AstroParticle Physics, The Ohio State University, 191 West Woodruff Avenue, Columbus, OH 43210, USA\label{aff53}
\and
Department of Physics, The Ohio State University, Columbus, OH 43210, USA\label{aff54}
\and
ESAC/ESA, Camino Bajo del Castillo, s/n., Urb. Villafranca del Castillo, 28692 Villanueva de la Ca\~nada, Madrid, Spain\label{aff55}
\and
Dipartimento di Fisica e Astronomia, Universit\`a di Bologna, Via Gobetti 93/2, 40129 Bologna, Italy\label{aff56}
\and
INAF-Osservatorio Astronomico di Padova, Via dell'Osservatorio 5, 35122 Padova, Italy\label{aff57}
\and
Universit\"ats-Sternwarte M\"unchen, Fakult\"at f\"ur Physik, Ludwig-Maximilians-Universit\"at M\"unchen, Scheinerstrasse 1, 81679 M\"unchen, Germany\label{aff58}
\and
Space Science Data Center, Italian Space Agency, via del Politecnico snc, 00133 Roma, Italy\label{aff59}
\and
Department of Physics "E. Pancini", University Federico II, Via Cinthia 6, 80126, Napoli, Italy\label{aff60}
\and
INAF-Osservatorio Astronomico di Capodimonte, Via Moiariello 16, 80131 Napoli, Italy\label{aff61}
\and
Instituto de Astrof\'isica e Ci\^encias do Espa\c{c}o, Universidade do Porto, CAUP, Rua das Estrelas, PT4150-762 Porto, Portugal\label{aff62}
\and
Faculdade de Ci\^encias da Universidade do Porto, Rua do Campo de Alegre, 4150-007 Porto, Portugal\label{aff63}
\and
Port d'Informaci\'{o} Cient\'{i}fica, Campus UAB, C. Albareda s/n, 08193 Bellaterra (Barcelona), Spain\label{aff64}
\and
INFN section of Naples, Via Cinthia 6, 80126, Napoli, Italy\label{aff65}
\and
Institute for Astronomy, University of Hawaii, 2680 Woodlawn Drive, Honolulu, HI 96822, USA\label{aff66}
\and
Dipartimento di Fisica e Astronomia "Augusto Righi" - Alma Mater Studiorum Universit\`a di Bologna, Viale Berti Pichat 6/2, 40127 Bologna, Italy\label{aff67}
\and
Instituto de Astrof\'{\i}sica de Canarias, V\'{\i}a L\'actea, 38205 La Laguna, Tenerife, Spain\label{aff68}
\and
European Space Agency/ESRIN, Largo Galileo Galilei 1, 00044 Frascati, Roma, Italy\label{aff69}
\and
Universit\'e Claude Bernard Lyon 1, CNRS/IN2P3, IP2I Lyon, UMR 5822, Villeurbanne, F-69100, France\label{aff70}
\and
UCB Lyon 1, CNRS/IN2P3, IUF, IP2I Lyon, 4 rue Enrico Fermi, 69622 Villeurbanne, France\label{aff71}
\and
Departamento de F\'isica, Faculdade de Ci\^encias, Universidade de Lisboa, Edif\'icio C8, Campo Grande, PT1749-016 Lisboa, Portugal\label{aff72}
\and
Instituto de Astrof\'isica e Ci\^encias do Espa\c{c}o, Faculdade de Ci\^encias, Universidade de Lisboa, Campo Grande, 1749-016 Lisboa, Portugal\label{aff73}
\and
Department of Astronomy, University of Geneva, ch. d'Ecogia 16, 1290 Versoix, Switzerland\label{aff74}
\and
Aix-Marseille Universit\'e, CNRS, CNES, LAM, Marseille, France\label{aff75}
\and
INAF-Istituto di Astrofisica e Planetologia Spaziali, via del Fosso del Cavaliere, 100, 00100 Roma, Italy\label{aff76}
\and
INFN-Padova, Via Marzolo 8, 35131 Padova, Italy\label{aff77}
\and
Aix-Marseille Universit\'e, CNRS/IN2P3, CPPM, Marseille, France\label{aff78}
\and
INFN-Bologna, Via Irnerio 46, 40126 Bologna, Italy\label{aff79}
\and
School of Physics, HH Wills Physics Laboratory, University of Bristol, Tyndall Avenue, Bristol, BS8 1TL, UK\label{aff80}
\and
Dipartimento di Fisica "Aldo Pontremoli", Universit\`a degli Studi di Milano, Via Celoria 16, 20133 Milano, Italy\label{aff81}
\and
INFN-Sezione di Milano, Via Celoria 16, 20133 Milano, Italy\label{aff82}
\and
Institute of Theoretical Astrophysics, University of Oslo, P.O. Box 1029 Blindern, 0315 Oslo, Norway\label{aff83}
\and
Jet Propulsion Laboratory, California Institute of Technology, 4800 Oak Grove Drive, Pasadena, CA, 91109, USA\label{aff84}
\and
Felix Hormuth Engineering, Goethestr. 17, 69181 Leimen, Germany\label{aff85}
\and
Technical University of Denmark, Elektrovej 327, 2800 Kgs. Lyngby, Denmark\label{aff86}
\and
Cosmic Dawn Center (DAWN), Denmark\label{aff87}
\and
Institut d'Astrophysique de Paris, UMR 7095, CNRS, and Sorbonne Universit\'e, 98 bis boulevard Arago, 75014 Paris, France\label{aff88}
\and
Max-Planck-Institut f\"ur Astronomie, K\"onigstuhl 17, 69117 Heidelberg, Germany\label{aff89}
\and
NASA Goddard Space Flight Center, Greenbelt, MD 20771, USA\label{aff90}
\and
Department of Physics and Astronomy, University College London, Gower Street, London WC1E 6BT, UK\label{aff91}
\and
Department of Physics and Helsinki Institute of Physics, Gustaf H\"allstr\"omin katu 2, 00014 University of Helsinki, Finland\label{aff92}
\and
Department of Physics, P.O. Box 64, 00014 University of Helsinki, Finland\label{aff93}
\and
Helsinki Institute of Physics, Gustaf H{\"a}llstr{\"o}min katu 2, University of Helsinki, Helsinki, Finland\label{aff94}
\and
Mullard Space Science Laboratory, University College London, Holmbury St Mary, Dorking, Surrey RH5 6NT, UK\label{aff95}
\and
SKA Observatory, Jodrell Bank, Lower Withington, Macclesfield, Cheshire SK11 9FT, UK\label{aff96}
\and
Centre de Calcul de l'IN2P3/CNRS, 21 avenue Pierre de Coubertin 69627 Villeurbanne Cedex, France\label{aff97}
\and
University of Applied Sciences and Arts of Northwestern Switzerland, School of Engineering, 5210 Windisch, Switzerland\label{aff98}
\and
Universit\"at Bonn, Argelander-Institut f\"ur Astronomie, Auf dem H\"ugel 71, 53121 Bonn, Germany\label{aff99}
\and
Department of Physics, Institute for Computational Cosmology, Durham University, South Road, Durham, DH1 3LE, UK\label{aff100}
\and
Universit\'e C\^{o}te d'Azur, Observatoire de la C\^{o}te d'Azur, CNRS, Laboratoire Lagrange, Bd de l'Observatoire, CS 34229, 06304 Nice cedex 4, France\label{aff101}
\and
Universit\'e Paris Cit\'e, CNRS, Astroparticule et Cosmologie, 75013 Paris, France\label{aff102}
\and
CNRS-UCB International Research Laboratory, Centre Pierre Bin\'etruy, IRL2007, CPB-IN2P3, Berkeley, USA\label{aff103}
\and
Institut d'Astrophysique de Paris, 98bis Boulevard Arago, 75014, Paris, France\label{aff104}
\and
Institute of Physics, Laboratory of Astrophysics, Ecole Polytechnique F\'ed\'erale de Lausanne (EPFL), Observatoire de Sauverny, 1290 Versoix, Switzerland\label{aff105}
\and
Aurora Technology for European Space Agency (ESA), Camino bajo del Castillo, s/n, Urbanizacion Villafranca del Castillo, Villanueva de la Ca\~nada, 28692 Madrid, Spain\label{aff106}
\and
Institut de F\'{i}sica d'Altes Energies (IFAE), The Barcelona Institute of Science and Technology, Campus UAB, 08193 Bellaterra (Barcelona), Spain\label{aff107}
\and
School of Mathematics, Statistics and Physics, Newcastle University, Herschel Building, Newcastle-upon-Tyne, NE1 7RU, UK\label{aff108}
\and
DARK, Niels Bohr Institute, University of Copenhagen, Jagtvej 155, 2200 Copenhagen, Denmark\label{aff109}
\and
Department of Physics and Astronomy, University of Waterloo, Waterloo, Ontario N2L 3G1, Canada\label{aff110}
\and
Perimeter Institute for Theoretical Physics, Waterloo, Ontario N2L 2Y5, Canada\label{aff111}
\and
Centre National d'Etudes Spatiales -- Centre spatial de Toulouse, 18 avenue Edouard Belin, 31401 Toulouse Cedex 9, France\label{aff112}
\and
Institute of Space Science, Str. Atomistilor, nr. 409 M\u{a}gurele, Ilfov, 077125, Romania\label{aff113}
\and
Consejo Superior de Investigaciones Cientificas, Calle Serrano 117, 28006 Madrid, Spain\label{aff114}
\and
Universidad de La Laguna, Departamento de Astrof\'{\i}sica, 38206 La Laguna, Tenerife, Spain\label{aff115}
\and
Dipartimento di Fisica e Astronomia "G. Galilei", Universit\`a di Padova, Via Marzolo 8, 35131 Padova, Italy\label{aff116}
\and
Universit\"at Innsbruck, Institut f\"ur Astro- und Teilchenphysik, Technikerstr. 25/8, 6020 Innsbruck, Austria\label{aff117}
\and
Satlantis, University Science Park, Sede Bld 48940, Leioa-Bilbao, Spain\label{aff118}
\and
Department of Physics, Royal Holloway, University of London, TW20 0EX, UK\label{aff119}
\and
Instituto de Astrof\'isica e Ci\^encias do Espa\c{c}o, Faculdade de Ci\^encias, Universidade de Lisboa, Tapada da Ajuda, 1349-018 Lisboa, Portugal\label{aff120}
\and
Cosmic Dawn Center (DAWN)\label{aff121}
\and
Niels Bohr Institute, University of Copenhagen, Jagtvej 128, 2200 Copenhagen, Denmark\label{aff122}
\and
Universidad Polit\'ecnica de Cartagena, Departamento de Electr\'onica y Tecnolog\'ia de Computadoras,  Plaza del Hospital 1, 30202 Cartagena, Spain\label{aff123}
\and
Kapteyn Astronomical Institute, University of Groningen, PO Box 800, 9700 AV Groningen, The Netherlands\label{aff124}
\and
Infrared Processing and Analysis Center, California Institute of Technology, Pasadena, CA 91125, USA\label{aff125}
\and
INAF, Istituto di Radioastronomia, Via Piero Gobetti 101, 40129 Bologna, Italy\label{aff126}
\and
Astronomical Observatory of the Autonomous Region of the Aosta Valley (OAVdA), Loc. Lignan 39, I-11020, Nus (Aosta Valley), Italy\label{aff127}
\and
INAF-Osservatorio Astronomico di Brera, Via Brera 28, 20122 Milano, Italy, and INFN-Sezione di Genova, Via Dodecaneso 33, 16146, Genova, Italy\label{aff128}
\and
ICL, Junia, Universit\'e Catholique de Lille, LITL, 59000 Lille, France\label{aff129}
\and
Instituto de F\'isica Te\'orica UAM-CSIC, Campus de Cantoblanco, 28049 Madrid, Spain\label{aff130}
\and
CERCA/ISO, Department of Physics, Case Western Reserve University, 10900 Euclid Avenue, Cleveland, OH 44106, USA\label{aff131}
\and
Technical University of Munich, TUM School of Natural Sciences, Physics Department, James-Franck-Str.~1, 85748 Garching, Germany\label{aff132}
\and
Max-Planck-Institut f\"ur Astrophysik, Karl-Schwarzschild-Str.~1, 85748 Garching, Germany\label{aff133}
\and
Donostia International Physics Center (DIPC), Paseo Manuel de Lardizabal, 4, 20018, Donostia-San Sebasti\'an, Guipuzkoa, Spain\label{aff134}
\and
IKERBASQUE, Basque Foundation for Science, 48013, Bilbao, Spain\label{aff135}
\and
Departamento de F{\'\i}sica Fundamental. Universidad de Salamanca. Plaza de la Merced s/n. 37008 Salamanca, Spain\label{aff136}
\and
Universit\'e de Strasbourg, CNRS, Observatoire astronomique de Strasbourg, UMR 7550, 67000 Strasbourg, France\label{aff137}
\and
Ludwig-Maximilians-University, Schellingstrasse 4, 80799 Munich, Germany\label{aff138}
\and
Max-Planck-Institut f\"ur Physik, Boltzmannstr. 8, 85748 Garching, Germany\label{aff139}
\and
Dipartimento di Fisica - Sezione di Astronomia, Universit\`a di Trieste, Via Tiepolo 11, 34131 Trieste, Italy\label{aff140}
\and
California Institute of Technology, 1200 E California Blvd, Pasadena, CA 91125, USA\label{aff141}
\and
Department of Physics \& Astronomy, University of California Irvine, Irvine CA 92697, USA\label{aff142}
\and
Department of Mathematics and Physics E. De Giorgi, University of Salento, Via per Arnesano, CP-I93, 73100, Lecce, Italy\label{aff143}
\and
INFN, Sezione di Lecce, Via per Arnesano, CP-193, 73100, Lecce, Italy\label{aff144}
\and
INAF-Sezione di Lecce, c/o Dipartimento Matematica e Fisica, Via per Arnesano, 73100, Lecce, Italy\label{aff145}
\and
Departamento F\'isica Aplicada, Universidad Polit\'ecnica de Cartagena, Campus Muralla del Mar, 30202 Cartagena, Murcia, Spain\label{aff146}
\and
Observatorio Nacional, Rua General Jose Cristino, 77-Bairro Imperial de Sao Cristovao, Rio de Janeiro, 20921-400, Brazil\label{aff147}
\and
CEA Saclay, DFR/IRFU, Service d'Astrophysique, Bat. 709, 91191 Gif-sur-Yvette, France\label{aff148}
\and
Department of Computer Science, Aalto University, PO Box 15400, Espoo, FI-00 076, Finland\label{aff149}
\and
Instituto de Astrof\'\i sica de Canarias, c/ Via Lactea s/n, La Laguna 38200, Spain. Departamento de Astrof\'\i sica de la Universidad de La Laguna, Avda. Francisco Sanchez, La Laguna, 38200, Spain\label{aff150}
\and
Caltech/IPAC, 1200 E. California Blvd., Pasadena, CA 91125, USA\label{aff151}
\and
Ruhr University Bochum, Faculty of Physics and Astronomy, Astronomical Institute (AIRUB), German Centre for Cosmological Lensing (GCCL), 44780 Bochum, Germany\label{aff152}
\and
Department of Physics and Astronomy, Vesilinnantie 5, 20014 University of Turku, Finland\label{aff153}
\and
Serco for European Space Agency (ESA), Camino bajo del Castillo, s/n, Urbanizacion Villafranca del Castillo, Villanueva de la Ca\~nada, 28692 Madrid, Spain\label{aff154}
\and
ARC Centre of Excellence for Dark Matter Particle Physics, Melbourne, Australia\label{aff155}
\and
Centre for Astrophysics \& Supercomputing, Swinburne University of Technology,  Hawthorn, Victoria 3122, Australia\label{aff156}
\and
Department of Physics and Astronomy, University of the Western Cape, Bellville, Cape Town, 7535, South Africa\label{aff157}
\and
Department of Astrophysics, University of Zurich, Winterthurerstrasse 190, 8057 Zurich, Switzerland\label{aff158}
\and
Department of Physics, Centre for Extragalactic Astronomy, Durham University, South Road, Durham, DH1 3LE, UK\label{aff159}
\and
IRFU, CEA, Universit\'e Paris-Saclay 91191 Gif-sur-Yvette Cedex, France\label{aff160}
\and
Oskar Klein Centre for Cosmoparticle Physics, Department of Physics, Stockholm University, Stockholm, SE-106 91, Sweden\label{aff161}
\and
Astrophysics Group, Blackett Laboratory, Imperial College London, London SW7 2AZ, UK\label{aff162}
\and
Univ. Grenoble Alpes, CNRS, Grenoble INP, LPSC-IN2P3, 53, Avenue des Martyrs, 38000, Grenoble, France\label{aff163}
\and
INAF-Osservatorio Astrofisico di Arcetri, Largo E. Fermi 5, 50125, Firenze, Italy\label{aff164}
\and
Dipartimento di Fisica, Sapienza Universit\`a di Roma, Piazzale Aldo Moro 2, 00185 Roma, Italy\label{aff165}
\and
Centro de Astrof\'{\i}sica da Universidade do Porto, Rua das Estrelas, 4150-762 Porto, Portugal\label{aff166}
\and
Dipartimento di Fisica, Universit\`a di Roma Tor Vergata, Via della Ricerca Scientifica 1, Roma, Italy\label{aff167}
\and
INFN, Sezione di Roma 2, Via della Ricerca Scientifica 1, Roma, Italy\label{aff168}
\and
HE Space for European Space Agency (ESA), Camino bajo del Castillo, s/n, Urbanizacion Villafranca del Castillo, Villanueva de la Ca\~nada, 28692 Madrid, Spain\label{aff169}
\and
Department of Astrophysical Sciences, Peyton Hall, Princeton University, Princeton, NJ 08544, USA\label{aff170}
\and
Theoretical astrophysics, Department of Physics and Astronomy, Uppsala University, Box 516, 751 37 Uppsala, Sweden\label{aff171}
\and
ASTRON, the Netherlands Institute for Radio Astronomy, Postbus 2, 7990 AA, Dwingeloo, The Netherlands\label{aff172}
\and
Anton Pannekoek Institute for Astronomy, University of Amsterdam, Postbus 94249, 1090 GE Amsterdam, The Netherlands\label{aff173}
\and
Center for Advanced Interdisciplinary Research, Ss. Cyril and Methodius University in Skopje, Skopje, Macedonia\label{aff174}
\and
Institute of Astronomy, University of Cambridge, Madingley Road, Cambridge CB3 0HA, UK\label{aff175}
\and
Space physics and astronomy research unit, University of Oulu, Pentti Kaiteran katu 1, FI-90014 Oulu, Finland\label{aff176}
\and
Institut de Physique Th\'eorique, CEA, CNRS, Universit\'e Paris-Saclay 91191 Gif-sur-Yvette Cedex, France\label{aff177}
\and
Center for Computational Astrophysics, Flatiron Institute, 162 5th Avenue, 10010, New York, NY, USA\label{aff178}}

 
\abstract
   {The \Euclid satellite will provide data on the clustering of galaxies and on the distortion of their measured shapes, which can be used to constrain
   and test the cosmological model.
   However, the increase in precision places strong requirements on the accuracy of the theoretical modelling for the observables and of the full analysis pipeline. In this paper, we investigate the accuracy of the calculations performed by the Cosmology Likelihood for Observables in Euclid (\CLOE), software able to handle both the modelling of observables and their fit against observational data for both the photometric and spectroscopic surveys of \Euclid, by comparing the output of \CLOE with external codes used as benchmarks. We perform such a comparison on the quantities entering the calculations of the observables, as well as on the final outputs of these calculations.
   Our results highlight the high accuracy of \CLOE when comparing its calculation against external codes for \Euclid\ observables on an extended range of operative cases. In particular, all the summary statistics of interest always differ less than $0.1\,\sigma$ from the chosen benchmark, and \CLOE predictions are statistically compatible with simulated data obtained from benchmark codes. The same holds for the comparison of the correlation function in configuration space for spectroscopic and photometric observables.
   }


\authorrunning{Euclid Collaboration: M. Martinelli et al.}

\titlerunning{CLOE. 4. Validation and performance}

\maketitle

\section{Introduction}

The current decade will see cosmological investigations profit from a significant increase in the precision of data, particularly those related to observations of the large-scale structure (LSS). Several galaxy surveys, such as \Euclid \citep{Laureijs11, EuclidSkyOverview}, the Vera C. Rubin Legacy Survey of Space and Time \citep[LSST;][]{Ivezic19}, and the {\it Nancy Grace Roman} Space Telescope \citep{Spergel15}, will observe the position and shape of billions of galaxies. This will allow us to obtain information on their clustering as well as on the lensing effect that intervening matter has on their shapes, thus effectively mapping the distribution of galaxies and matter across the Universe.

The main goal of such observations is to shed light on the key open questions in the cosmological community, such as understanding the mechanism responsible for the late-time acceleration of the expansion of the Universe \citep{RieFilCha1998, PerAldGol1999}. Current models range from those that require the presence of a scalar field induced by a dark energy component \citep{PeeRat2003} to others that instead invoke a modification of the laws of general relativity on cosmological scales \citep{CliFerPad2012}. At the same time, other fundamental questions and goals are related to the nature of dark matter \citep{Feng2010}, the measurement of the mass of neutrinos via their impact on the LSS \citep{LesPas2006}, as well as the potential detection of primordial non-Gaussianities in the epoch immediately after inflation \citep{DesSel2010}. All of these components leave distinctive imprints on the LSS observables, since they can strongly affect the large-scale matter and galaxy distribution expected from the standard model of cosmology, based on cold pressureless dark matter (CDM) and on a cosmological constant, $\Lambda$, the \LCDM\ model.

The \Euclid mission \citep{Laureijs11} was specifically designed to tackle these open questions and is expected to deliver a significant improvement in our knowledge in these fields \citep{EuclidSkyOverview}. 
However, in order to achieve these goals, the exquisite precision of the measurements that will be collected by Stage-IV surveys demands a corresponding level of accuracy in the theoretical modelling of the LSS observables. 

For this reason, the Euclid Collaboration has decided to dedicate a major effort to the development of software to compute theoretical predictions for the main scientific observables: the Cosmology Likelihood for Observables in Euclid (\CLOE). 
This code implements the modelling of the primary photometric and spectroscopic observables that will be used by \Euclid, following the theoretical recipe described in 
\citet{EP-CLOE1}, hereafter \citetalias{EP-CLOE1}. Building a standalone likelihood and theory code for \Euclid is essential, as we aim to establish a homogeneous and flexible framework that consistently models the primary probes of the mission --  weak lensing and galaxy clustering -- within a unified approach. Furthermore, the decision to build a new code for the analysis of \Euclid data allows us to improve the modelling and analysis of systematic effects. Indeed, \CLOE includes nuisance parameters that capture effects that could be neglected in previous analyses, given the lower sensitivity with respect to \Euclid. Moreover, it is fully written in \python, while still achieving the performance required by a likelihood analysis. More information on these new implementations is provided in \citetalias{EP-CLOE1} and \citet{Paper2}, hereafter \citetalias{Paper2}.

At the practical level, \CLOE obtains theoretical predictions given a set of cosmological parameters using an interface with \cobaya\footnote{\url{https://cobaya.readthedocs.io/en/latest/}} \citep{Torrado:2020dgo}, which can retrieve the main ingredients for our theoretical expectations from publicly available Boltzmann solvers, \camb\footnote{\url{https://camb.readthedocs.io/en/latest/}} \citep{Lewis:1999bs,Howlett:2012mh} and \class\footnote{\url{http://class-code.net/}} \citep{2011JCAP...07..034B}. The structure of \CLOE, its interface with \cobaya, and the way theoretical calculations are performed are presented in detail in 
\citetalias{Paper2}.

In addition to computation of cosmological observables, the interface with \cobaya allows the user to employ \CLOE as a likelihood code, which means comparing theoretical predictions with observational data. By sampling the parameter space and performing this comparison iteratively, \CLOE allows us to infer the probability distribution of the free parameters of the model under analysis. This procedure is described in 
\citet{EP-ClOE3}, hereafter \citetalias{EP-ClOE3},  
where \CLOE is used to derive forecasts on the parameters of the \LCDM model and some of its extensions (e.g. $w_0w_{\rm a}$CDM, non-flat cosmologies) from synthetic \Euclid data using a realistic set of Markov chain Monte Carlo (MCMC) runs. 

The increased precision of the observations that we expect from \Euclid\ requires the inclusion of several systematic effects in the modelling of the observables, as neglecting these can produce significant biases in the final results obtained while analysing the data \citep{EUCLID:2020jwq,Euclid:2020tff}. Given the complexity of some of the calculations required, it is crucial to assess the accuracy of \CLOE,  to ensure the quality of the computed theoretical predictions. The choices made in performing the calculations, such as the numerical integrations or the sampling of cosmological functions in redshift and scale, can significantly affect the output of the code and, consequently, the results of the analysis. For this reason, in this paper we aim to compare the calculations performed by \CLOE against independent benchmarks, assessing the reliability of this software. This effort is similar to the benchmarking approach taken by other collaborations for codes with similar purposes, such as the Core Cosmology Library (\ccl\footnote{\url{https://github.com/LSSTDESC/CCL/}}, \citealt{pyccl2019}) developed for LSST \citep{LSSTScience:2009jmu,LSSTDarkEnergyScience:2018jkl}. Although in this paper our aim is to validate the accuracy of our calculations, the impact of the systematic effects modelled within \CLOE will be investigated in a subsequent publication \citep{EP-CLOE6}.

To reach our goal, we assess the accuracy of the theoretical calculations performed by \CLOE, benchmarking them against the results of external codes. We quantify the agreement in units of the observational error, thus ensuring that any difference found is much lower than the statistical uncertainty and, therefore, that it will not impact the real-data analysis. While in this work we provide an assessment of the accuracy of \CLOE calculations, an extensive investigation of the methodology used to obtain the observables, their uncertainty, together with a comparison between different non-linear recipes, will be presented in Euclid Collaboration: Crocce et al. (in prep.), Euclid Collaboration: Carrilho et al. (in prep.), Euclid Collaboration: Moretti et al. (in prep.), and Euclid Collaboration: Sciotti et al. (in prep.).

This paper is structured as follows. In Sect.\ \ref{sec:benchcodes} we describe the external codes used to obtain the benchmarks on the observables of interest, while in Sect.\ \ref{sec:settings} we provide details on the settings used for the validation, describing the different effects taken into account, as well as the cosmological assumptions. In Sect.\ \ref{sec:methods}, we outline the methods used to validate \CLOE against independent codes, and we define the metrics used to assess their agreement. Section \ref{sec:validation} contains the validation results for the power spectra observables related to the spectroscopic and photometric observables surveys, including the intermediate quantities that are used to perform these calculations, while Section \ref{sec:corrfuncs} validates the computation of real-space correlation functions.
We draw our conclusions in Sect.\ \ref{sec:conclusions}.

\section{External codes and setup for benchmark generation}\label{sec:benchcodes}

\begin{table}
\begin{center}
\caption{Summary of connections between equations in \citetalias{EP-CLOE1} and \CLOE quantities discussed in this work.}
\resizebox{\columnwidth}{!}{
\begin{tabular}{l|l|l}
\hline
Quantity & Expression & Equation \\
\hline
Weak lensing power spectrum                     & $C_{ij}^{\rm LL}(\ell)$               & (32)\\
Photometric galaxy power spectrum               & $C_{ij}^{\rm GG}(\ell)$               & (40)\\
Galaxy-galaxy lensing power spectrum            & $C_{ij}^{\rm GL}(\ell)$               & (54)\\
Weak lensing correlation function               & $\xi_{ij}^\pm(\theta)$                & (70) \\
Photometric galaxy correlation function         & $\xi_{ij}^{\rm GG}(\theta)$           & (73) \\
Galaxy-galaxy lensing correlation function      & $\xi_{ij}^{\rm GL}(\theta)$           & (74) \\
Power spectrum Legendre multipoles              & $P_\ell(k, z)$                        & (90) \\
AP-distorted power spectrum Legendre multipoles & $P_{\rm obs, \ell}(k^{\rm fid}, z)$   & (94) \\
Correlation function Legendre multipoles        & $\xi_{{\rm obs},\ell}(s^{\rm fid},z)$ & (104) \\
\hline
\end{tabular}
}
\label{tab:func_eqs}
\end{center}
\end{table}

As the purpose of this paper is to validate the predictions obtained with \CLOE, we compare its results with those of external codes that have previously been validated, at least in a subset of the relevant cases. This section provides a description of these external codes, along with their corresponding references, and outlines the setup adopted to generate the benchmark predictions used for comparison throughout the paper.

In this work, we do not define all the observables that are compared with external codes. The definition of these can be found in \citetalias{EP-CLOE1}, where the intermediate quantities used for the calculations are also defined. We report in \cref{tab:func_eqs} the reference for the equations of the observables that we discuss to facilitate their retrieval in \citetalias{EP-CLOE1}.

\subsection{External benchmark for 3\texorpdfstring{$\times$}{X}2pt}

The main observable of the photometric survey of \Euclid is the set of angular power spectra $\CABij(\ell)$, defined in \citetalias{EP-CLOE1}, for weak lensing (WL), galaxy clustering, and their cross-correlation. These spectra are obtained from \texttt{CLOE} for all the redshift bin combinations given a binned galaxy distribution. We compute $\CABij(\ell)$ in 270 multipole bins, logarithmically spaced.
Similarly, we consider the projection of power spectra in the two points correlation functions (2PCF) $\xiGGij(\theta)$, $\ximinusij(\theta)$, $\xiplusij(\theta)$, and $\xiGLij(\theta)$, also defined in \citetalias{EP-CLOE1}. These correlation functions are computed in 40 logarithmically spaced bins in the angle $\theta$, contained in the interval $\theta\in\left[0.1,10\right]\,\deg$.

For what concerns the predictions of \CLOE for photometric observables, we rely on two different codes to produce benchmarks: \life and \ccl. While \ccl is a code that has already been extensively used by the cosmological community, this is not the case for \life, which is a private code developed to obtain predictions for the \Euclid\ survey.

However, due to differences in the assumed recipe, we cannot compare \CLOE with \ccl for all the cases detailed in Sect.\ \ref{sec:settings}. Therefore, we choose to present most of our validation results focusing on \life, while we show the outcome of the comparison with \ccl, for the cases where this is possible, in Appendix~\ref{sec:pyccl_comp}.

\subsubsection{\life}

\life (Likelihood and Forecasts for Euclid) is a collection of codes written in {\tt Mathematica}\footnote{Courtesy of V. F. Cardone.} to perform the computation of the 3\texttimes2pt observables for likelihood evaluation and Fisher matrix forecasts. It has been validated against the code implementation adopted during the Fisher forecast analysis from \citet{Blanchard-EP7}, checking that, for the same settings, it provides the same estimates of the marginalised constraints on the cosmological parameters, also comparing the intermediate quantities used in calculations. Although not optimised and much slower than any \python implementation, \life has been designed with the validation of \CLOE in mind. As a consequence, the same exact recipe is also implemented with the possibility of turning on and off different contributions to compare against other codes. For the same reason, it does not compute the matter power spectrum internally, but imports it as an external quantity (to be interpolated) so that one can be sure that any difference with \CLOE is not due to errors in this critical quantity.  However, the fact that it is coded in a completely different language (with its own interpolation and integration routines) represents a further benefit, allowing us to verify that, for the same input and the same recipe, two independent implementations yield compatible quantities, which is what we want to verify in this work. 

\life can also be used to perform Fisher matrix forecasts for \Euclid using the same assumptions (in terms of cosmological model) and settings (e.g. redshift distributions, number of redshift bins, and angular multipoles), adopting a semi-analytic method to compute the derivatives of the observables. Although it is well known that the Fisher matrix method provides lower limits on the marginalised constraints on the cosmological parameters under the assumption of Gaussian likelihood, these forecasts can be used to check that the results from inference are in the same ballpark 
\citepalias[see][]{EP-ClOE3}.

\subsubsection{\ccl}\label{sec:CCL}
\ccl is a modern cosmological library designed to compute a large number of cosmological observables for a variety of cosmological models. These range from background quantities, power spectra, and correlation functions to halo model ingredients; it is interfaced with both \camb\ -- the default choice -- and \class, and it has been extensively validated against existing codes \citep{pyccl2019}. The core of \ccl is developed in \texttt{C}, allowing it to achieve high performance. However, a thoroughly documented \python interface, \pyccl,\footnote{\texttt{\url{https://ccl.readthedocs.io/en/latest/index.html}}} is also available, combining efficiency and user-friendliness in this way.

For the present exercise, we used the third release of the code (more specifically, \texttt{v3.0.2}).
The theory predictions are computed using as input the cosmological parameters, redshift distributions, and systematics defined in Sect.\ \ref{sec:photo_settings}, as well as the matter power spectrum extracted from \CLOE; the growth factor is instead obtained from \ccl itself. The harmonic-space power spectra are computed in the Limber approximation via the \texttt{QAG} adaptive integration method of \texttt{GSL} (GNU Scientific Library,\footnote{\texttt{\url{http://www.gnu.org/software/gsl/}}} \citealt{galassi2018GNU}); these are then transformed into the 2PCF via the FFTlog method. 

The \ccl routines can produce all the cases listed in \cref{tab:cases}, except for the ones including redshift space distortions (RSD) (i.e. P23 and from P11 to P22), for which the implemented recipe differs from the one used in \CLOE, which, as outlined in 
\citetalias{EP-CLOE1}, 
follows \citet{Kostas2019}. For these cases, only the comparison with \life will be shown. The results of the comparison with \ccl can be found in Appendix \ref{sec:pyccl_comp}.

\subsection{External benchmark for spectroscopic galaxy clustering (GCsp)}
\label{sec:gcspec_bench}

The main observable of the spectroscopic survey of \Euclid is the set of Legendre multipoles of the anisotropic galaxy power spectrum, $P_{\mathrm{obs}, \ell}(k^{\mathrm{fid}}, z)$, as defined in \citetalias{EP-CLOE1} and Euclid Collaboration: Moretti et al. (in prep.). These multipoles are obtained by projecting the two-dimensional anisotropic power spectrum onto the Legendre polynomials $\mathcal{L}_\ell(\mu)$. Under the assumptions of linear theory, only the first three even multipoles are non-vanishing, corresponding to $\ell \in \{0, 2, 4\}$.\footnote{Non-linear corrections can source non-zero even multipoles with order beyond $\ell = 4$. However, due to their large statistical uncertainties, these additional terms are completely sub-dominant in terms of constraining power and do not carry significant information with respect to the combination $\ell\in\{0,2,4\}$.} Similarly, in configuration space, we consider an equivalent expansion over the Legendre polynomials, leading to the 2PCF Legendre multipoles, $\xi_{{\rm obs},\ell}(s^{\rm fid},z)$.

For the purpose of the validation carried out here, we generate external benchmark predictions without including geometrical distortions caused by adopting a fiducial cosmology that differs from the true one. In other words, we neglect the Alcock–Paczynski \citep[AP;][]{AlcPac1979} effect, which would rescale the transverse, $k_\perp$, and line-of-sight, $k_\parallel$, components of the wavevector $k$. As a result, the effective quantities we compare are the Legendre multipoles of the power spectrum $P_\ell(k, z)$, computed without AP corrections. Since AP effects depend solely on background quantities, which are validated separately in Appendix~\ref{sec:class_comp}, we do not explore their impact further in this work.

In all cases defined in Sect. \ref{sec:spectro_settings} for the power spectrum Legendre multipoles, the data vectors and covariance matrices are computed at the mean redshift of four distinct redshift bins, namely $\{(0.9, 1.1),\ (1.1, 1.3),\ (1.3, 1.5),\ (1.5, 1.8)\}$, and are sampled using 350 linear $k$ bins in the range $(0.001,\, 0.35)\,\mathrm{Mpc}^{-1}$. In contrast,  for validation carried out in the configuration space, the 2PCF is sampled with linear bins of $\Delta s=5\,{\rm Mpc}$ over the range $(37.5, 387.5)\,{\rm Mpc}$.

The external benchmarks are produced using two different codes, \pbj and \coffe. While the former adopts the same non-linear modelling recipe as \CLOE in Fourier space, the latter offers a convenient framework to test the pipeline that converts Legendre multipoles from Fourier to configuration space. In the following sections, we briefly summarise the main features of the two codes.

\subsubsection{\pbj}

The implementation of the non-linear model in \CLOE is validated against that in the \texttt{Power spectrum Bispectrum Joint} code (\pbj),\footnote{Courtesy of C. Moretti, E. Sefusatti, A. Oddo.} a non-public \python package that has been employed in several analyses involving the full shape of the galaxy power spectrum, $\Pgg(k)$, and bispectrum, $\Bggg(\vec{k}_1,\vec{k}_2)$. These include both comparisons to numerical simulations in the context of model selection \citep{Oddo2020, Oddo2021}, as well as cosmological inference from real data, such as the recent analysis of the BOSS DR12 sample (Euclid Collaboration: Moretti et al. in prep.).

The code features an implementation of the widely adopted effective field theory of large-scale structure (EFTofLSS) model, based on the same theoretical prescription presented in \citet{EP-Pezzotta}, which is precisely what has been reproduced in the current version of \CLOE. A common feature of models based on non-linear perturbation theory and the EFTofLSS framework is that non-linear corrections to the galaxy (and matter) power spectrum arise from the coupling of different physical scales. This coupling is mediated by kernel functions that encapsulate the physics of gravitational instability, galaxy bias, and RSD. Specifically, the non-linear correction $\Delta P$ at a given wavenumber $k$ can be expressed as a convolution of the power spectrum with itself,
\be
    \Delta P(k) = \int \frac{\diff^3 q}{\left(2\pi\right)^3} \, K(\vec{q}, \vec{k}-\vec{q}) \, P(\vec{q}) \, P(\left|\vec{k}-\vec{q}\right|)\,,
\ee
where the kernel $K(\vec{q}_1,\vec{q}_2)$ captures the non-linear mode coupling between the wavemodes $\vec{q}_1$ and $\vec{q}_2$. To efficiently evaluate these integrals, \pbj relies on the \texttt{FASTPT} library \citep{McEwen2016}, which takes advantage of the possibility of expanding $K(\vec{q}_1,\vec{q}_2)$ in spherical harmonics, enabling the use of fast Fourier transform techniques for rapid and accurate computation.

The implementation of spectroscopic observables in \CLOE follows the same approach as \pbj, and both codes adopt identical parametrisation for the EFTofLSS nuisance parameters. This consistency allows for a direct one-to-one comparison of each validation case described in Sect.~\ref{sec:spectro_settings}, without the need to translate between parameter bases. As a result, we can robustly validate the \CLOE predictions for the galaxy power spectrum multipoles $P_\ell(k)$ across all 20 benchmark scenarios presented in Sect.~\ref{sec:spectro_settings}.

\subsubsection{\coffe}
\label{sec:coffe}

Validation of the Legendre multipoles is performed using predictions from the \texttt{COrrelation Function Full-sky Estimator} (\coffe) code\footnote{\texttt{\url{https://github.com/JCGoran/coffe}}}, a public package designed to efficiently compute two-point statistics in configuration space under linear theory assumptions. In addition to the standard RSD signal, \coffe offers a rigorous treatment of several relativistic effects, including wide-angle corrections and magnification bias. We refer the reader to the \coffe-related literature for further details: theoretical modelling of relativistic effects in the 2PCF is presented in \citet{TanBonDur2018}, their implementation in \citet{TanJelBon2018}, and an assessment of the flat-sky approximation in \citet{Jelic-Cizmek_2021}.

In this work, we employ \coffe to validate the accuracy of the Fourier-to-configuration space mapping used in \CLOE, which is based on the numerical evaluation of a Hankel transform. For this purpose, we consider a single benchmark case based on a \LCDM cosmology with linear-theory predictions, incorporating linear bias and the leading-order RSD effect describing large-scale infall due to galaxy peculiar velocities \citep{Kaiser1987}.

\section{Validation settings}\label{sec:settings}

One of the main purposes of this validation is to assess the reliability of \CLOE when different effects are included in the modelling of \Euclid observables. For this reason, we compare the predictions of \CLOE against other benchmark codes, exploring several combinations of the options available in \CLOE. Although we do not perform a full exploration of the parameter space,  as would be possible with a Monte Carlo approach, the selected cases are sufficient to validate the main features and systematic effects included in \CLOE.

All results presented in this paper were obtained using version \texttt{v2.0.2} of \CLOE,\footnote{The theoretical predictions were also benchmarked for v2.1, finding compatible results.} the same version used to derive forecasts in \citetalias{EP-ClOE3}. The theoretical framework and implementation details of this version are described in \citetalias{EP-CLOE1} and \citetalias{Paper2}, respectively.

\subsection{Photometric settings}\label{sec:photo_settings}

For the validation of the photometric observables, we report 23 cases in \cref{tab:cases} (labelled P01, P02, ..., P23) for which we run our validation pipeline. Switching between these cases, we vary some of the settings available in \CLOE to compute the observables. In particular, we focus on variations related to the following quantities.
\begin{itemize}
    \item[(i)] Cosmology: we choose ten different cosmologies, described in \cref{tab:cosmopars}, where we report the current values ($z=0$) of the energy density parameters for matter $\Omega_{\rm m}$, dark energy $\Omega_{\rm DE}$, and baryons $\Omega_{\rm b}$, together with the reduced Hubble parameter $h$, the tilt of the primordial power spectrum $n_{\rm s}$, and the perturbation amplitude parameter $\sigma_8$. We include  \LCDM extensions such as evolving dark energy equation of state -- described by the $w_0$ and $w_a$ parameters -- and non-flat cosmologies -- describing scenarios with $\Omega_{\rm m}+\Omega_{\rm DE}\neq1$.
    \item[(ii)] Galaxy distribution $n(z)$: we choose different galaxy redshift distributions, matching the one used in \citet{Blanchard-EP7}, labelled Inter Science working group Taskforce Forecast (ISTF), the one described in Sect.\ 5.1 of 
    \citetalias{EP-CLOE1}, labelled Science Performance Verification 3 (SPV3), and an intermediate one (Science Performance Verification 2, SPV2), which still contains 13 bins as in SPV3, but with more irregular bin distributions. These choices allow us to test the stability of the code moving from smooth, simple $n(z)$ (ISTF) to more complicated and realistic distributions, investigating potential numerical issues.
    \item[(iii)] Intrinsic alignment (IA): we choose four cases for this systematic effect, regulated by the three IA parameters ($A_{\rm IA},\eta_{\rm IA},\beta_{\rm IA}$). These are cNLA ($0.37,0.00,0.00$), zNLA ($0.16,1.66,0.00$), eNLA ($1.720,-0.41,2.17$), and a simple noIA case ($0,0,0$) in which IA is completely neglected 
    \citepalias[see][and references therein]{EP-CLOE1}. Choosing these different settings allows us to test \CLOE\ in simple scenarios where this systematic effect is absent, as well as different degrees of complexity, by generalising their modelling.
    \item[(iv)] Galaxy bias $b_{\rm g}(z)$: we consider both a binned approach, with one parameter per redshift bin, and a third-order polynomial expression to describe the $b_{\rm g}(z)$ function. For the former case, we consider an `unbiased' case, with all parameters vanishing, a `linint' case, where $b_{\rm g}(z)$ is obtained as the linear interpolation of the binned parameters, and a `constant' case, where the bias is assumed to be constant in each redshift bin. The two modelling methods tested with these choices provide a stress-test for the reliability of \CLOE\ comparing its performances when switching from a smooth polynomial function to the more agnostic binned approach.
    \item[(v)] Magnification bias: as for the galaxy bias, we test our calculation modelling the magnification bias using the same approaches as for $b_{\rm g}(z)$.
    \item[(vi)] RSD: we compare \CLOE calculations with the external benchmark in the two available settings for the large-scale RSD effect, that is, when this is considered and when it is neglected. This allows us to test the correct implementation of the RSD model, described in \citet{Tanidis-TBD}.
\end{itemize}
Further details on the modelling of the nuisance parameters can be found in 
\citetalias{EP-CLOE1}, 
while their impact on the cosmological constraints will be discussed in \citet{EP-CLOE6} also through a more thorough exploration of their parameter space.

Of the cases listed in \cref{tab:cases}, our reference case is P23, which is also based on the settings used to generate synthetic data to obtain forecast results 
\citepalias{EP-ClOE3}.
Although for all cases, we show the summary validation values for all quantities (see Sect.\ \ref{sec:methods}), we will highlight in Appendix \ref{sec:reference} more details for our reference, also showing the trends of the differences for each of the validated quantities.

\begin{table}
\centering
\caption{Validation cases in which the \CLOE calculation of \Euclid photometric observables are compared with external software.}
\resizebox{\columnwidth}{!}{
\begin{tabular}{ccccccc} 
\hline
Id & Cosmology & $n(z)$ & IA & galaxy bias & magnification bias & RSD\\
 \hline 
P01 & F1 & ISTF & noIA & linint     & unbiased   & no  \\ 
P02 & F1 & ISTF & cNLA & linint     & unbiased   & no  \\ 
P03 & F1 & ISTF & zNLA & linint     & unbiased   & no  \\ 
P04 & F1 & ISTF & eNLA & linint     & unbiased   & no  \\ 
P05 & F1 & ISTF & eNLA & unbiased   & unbiased   & no  \\ 
P06 & F1 & ISTF & eNLA & constant   & unbiased   & no  \\ 
P07 & F1 & SPV2 & eNLA & constant   & unbiased   & no  \\ 
P08 & F1 & SPV3 & eNLA & constant   & unbiased   & no  \\ 
P09 & F1 & SPV3 & eNLA & constant   & linint     & no  \\ 
P10 & F1 & SPV3 & eNLA & constant   & constant   & no  \\ 
P11 & F1 & SPV3 & eNLA & constant   & unbiased   & yes \\ 
P12 & F1 & SPV3 & eNLA & constant   & linint     & yes \\ 
P13 & F1 & SPV3 & eNLA & constant   & constant   & yes \\ 
P14 & F2 & SPV3 & eNLA & constant   & constant   & yes \\ 
P15 & F3 & SPV3 & eNLA & constant   & constant   & yes \\ 
P16 & F4 & SPV3 & eNLA & constant   & constant   & yes \\ 
P17 & F5 & SPV3 & eNLA & constant   & constant   & yes \\ 
P18 & F6 & SPV3 & eNLA & constant   & constant   & yes \\ 
P19 & F7 & SPV3 & eNLA & constant   & constant   & yes \\ 
P20 & F8 & SPV3 & eNLA & constant   & constant   & yes \\ 
P21 & F9 & SPV3 & eNLA & constant   & constant   & yes \\ 
P22 & F10 & SPV3 & eNLA & constant   & constant   & yes \\
P23 & F1 & SPV3 & zNLA & polynomial & polynomial & yes \\ 
\hline
\end{tabular} 
}
\label{tab:cases}
\end{table}

\begin{table}
\centering
\caption{Values of the cosmological parameters in ten cosmologies considered for the validation of cosmological quantities and the main photometric and spectroscopic observables.}
\resizebox{\columnwidth}{!}{
\begin{tabular}{lcccccccc} 
\hline
Id & $\Omega_{\rm m}$ & $\Omega_{\rm DE}$ & $\Omega_{\rm b}$ & $w_{0}$ & $w_{a}$ & $h$ & $n_{\rm s}$ & $\sigma_{8}$\\
 \hline 
F1 & 0.320 & 0.680 & 0.050 & $-1.000$ & 0.000  & 0.6737 & 0.9660 & 0.8155\\
F2 & 0.317 & 0.676 & 0.051 & $-1.017$ & 0.010  & 0.6775 & 0.9653 & 0.8222\\
F3 & 0.327 & 0.670 & 0.048 & $-0.913$ & $-0.420$ & 0.6704 & 0.9561 & 0.8095\\
F4 & 0.316 & 0.689 & 0.046 & $-1.066$ & 0.313  & 0.6616 & 0.9524 & 0.8199\\
F5 & 0.321 & 0.724 & 0.051 & $-0.914$ & $-0.050$ & 0.6729 & 0.9643 & 0.7996\\
F6 & 0.306 & 0.702 & 0.047 & $-1.039$ & 0.046  & 0.6730 & 0.9699 & 0.8248\\
F7 & 0.326 & 0.721 & 0.050 & $-0.985$ & 0.428  & 0.6554 & 0.9461 & 0.8059\\
F8 & 0.307 & 0.664 & 0.045 & $-1.116$ & 0.268  & 0.6724 & 0.9708 & 0.8273\\
F9 & 0.337 & 0.660 & 0.053 & $-0.873$ & $-0.399$ & 0.6658 & 0.9383 & 0.8099\\
F10 & 0.334 & 0.630 & 0.057 & $-0.903$ & $-0.587$ & 0.6720 & 0.9779 & 0.8057\\
\hline
\end{tabular} 
}
\label{tab:cosmopars}
\end{table}

\subsection{Spectroscopic settings}
\label{sec:spectro_settings}

In terms of the spectroscopic probe, we identified 20 different cases for the validation, labelled as S01, S02, ..., S20. The first ten cases assume a fixed cosmology (F1 in \cref{tab:cosmopars}) and vary the nuisance parameters,\footnote{The selected values of the nuisance parameters have been randomly drawn from the marginalised posterior distribution obtained from a specific configuration of the chains run in \citetalias{EP-ClOE3}.} whereas the last ten cases keep the latter fixed and explore the ten different cosmologies shown in \cref{tab:cosmopars}. 
In terms of nuisance parameters of EFTofLSS, we assume a minimal configuration corresponding to that adopted to obtain MCMC forecasts in \citetalias{EP-ClOE3}. This includes six non-zero nuisance parameters, corresponding to the linear $b_1$ and quadratic bias $b_2$, three EFTofLSS counterterms $(c_0,\,c_2,\,c_4)$ mostly meant to describe the small-scale RSD damping induced by the Fingers of God, and a shot-noise parameter $\alpha_{P}$ that models constant offsets from Poisson predictions, that is, $1/\overline{n}$ \citep[][Euclid Collaboration: Crocce et al. in prep., Euclid Collaboration: Moretti et al. in prep.]{IvaSimZal2020}.

On the other hand, the validation of 2PCF is carried out assuming a single configuration based on linear theory, as described in Sect. \ref{sec:coffe}. In this case, we assume a \LCDM cosmology for the comparison. Specifically, the fiducial parameters are $\Omega_{\rm m}=0.319$, $\Omega_{\rm DE}=0.681$, $\Omega_{\rm b}=0.049$, $h=0.67$, $n_{\rm s}=0.96$, and $\sigma_{8}=0.83$. 

\section{Validation methods}\label{sec:methods}

The main purpose of this paper is to validate the theoretical predictions for \Euclid observables obtained by \CLOE against external benchmarks. While we are mostly interested in validating the final results, we also examine the intermediate quantities, which are the building blocks of the observables calculation. 
In this section, we present the different methods used to estimate the agreement between the quantities predicted by \CLOE and the corresponding ones from the codes we use as a benchmark. Similarly to what is done for other public codes, such as \texttt{CCL} \citep[see][]{pyccl2019}, we employ different metrics depending on whether or not the quantity under examination is actually observable.

\subsection{Symmetric mean absolute percentage error}\label{sec:SMAPE}

When assessing the accuracy in our calculations for quantities that are not observables, thus lacking an observational error, we estimate their agreement with the benchmark using the symmetric mean absolute percentage error (SMAPE)
\begin{equation}\label{eq:SMAPE}
    {\rm SMAPE} = \frac{100}{N}\sum_{i=1}^N{\frac{\left|F^{\rm \CLOE}_i-F^{\rm Bench}_i\right|}{\left(\left|F^{\rm \CLOE}_i\right|+\left|F^{\rm Bench}_i\right|\right)/2}}\,,
\end{equation}
where $F^{\rm \CLOE}_i$ are the quantities computed by \CLOE, $F^{\rm Bench}_i$ is the corresponding benchmark quantity, and the index $i$ runs over the $N$ available elements of the (vectorised) arrays.

The SMAPE estimator has the advantage of quantifying the agreement between two functions relative to the average between the compared quantities, thus without assuming which of the two is correct. Moreover, the SMAPE also allows us to obtain a single value for each of the cases in which we perform our validation (see Sect.\ \ref{sec:settings}), thus returning a unique estimation of agreement for each case.

The disadvantage of using this approach is that it does not allow us to observe trends in relation to redshift or scale differences. To address this issue, we also computed the value of the SMAPE at each point in the data vectors and plotted it as a function of redshift or scale, depending on the quantity of interest.

\subsection{Comparison of observable quantities}\label{sec:err_comparison}

While we rely on the SMAPE to assess the difference against benchmark calculations for most quantities, this is not the case for the final observables. For these, we can estimate the observational uncertainty with which \Euclid can obtain measurements. We choose to estimate our accuracy in terms of this uncertainty.

We define a notion of distance for a given observable $O$ as
\begin{equation}\label{eq:errdist}
    d_i = \frac{\left|O^{\,\rm \CLOE}_i-O^{\,\rm Bench}_i\right|}{\sigma_{O_i}}\,,
\end{equation}
where the index $i$ runs over all the available elements of the observable array, and $\sigma_{O_i}$ is the observational error associated with each element, obtained using $O^{\,\rm Bench}_i$ as a reference to compute the error. 

The distance defined in \cref{eq:errdist} allows us to visualise the difference trend by computing it for each element of the observable array. However, we also want to obtain a single value to quantify the agreement with the benchmark for each validation case. Therefore, we also compute an average and a maximum distance over each observable array, defined as
\begin{align}
    d_{\rm mean} &= \frac{1}{N}\sum_{i=1}^N{d_i
    }\,,\label{eq:mean_errdist}\\
    d_{\rm max} &= \max_id_i
    \,.\label{eq:max_errdist}
\end{align}
With these definitions, it is possible to quantify the agreement between the observables computed by \CLOE and the benchmark in terms of the \Euclid precision. This allows us to define an accuracy threshold above which a difference between our calculation and the benchmark would be too significant to neglect. We set such a threshold to be $d_{\rm max}<0.1$, defining all quantities for which the distance is within $10\%$ of the observational uncertainty as validated \citep{Massey:2012cz,Euclid:2020tff}.

 Although the distance defined in \cref{eq:errdist} provides an intuitive quantification of the agreement between \texttt{CLOE} and the benchmark codes, and is widely used in the literature \citep[][]{pyccl2019,LSSTDarkEnergyScience:2022lno}, it has two main drawbacks:
\begin{itemize}
    \item it does not provide an overall estimate of the statistical compatibility between the prediction of \CLOE and those of the benchmark;
    \item it does not take into account the full observational covariance, since it assumes no correlation between different data bins.
\end{itemize}

\noindent Therefore, we include an additional quantification of the agreement between the codes in terms of $\chi^2$, which allows us to solve these issues.

In order to assess whether the observables computed by \texttt{CLOE} are in agreement with the benchmark within \Euclid\ specifications, we compute the reduced $\chi^2$, that is, the $\chi^2$ divided by the number of degrees of freedom $N_{\rm dof}$, obtained by comparing \texttt{CLOE} predictions to those of the benchmark as
\begin{equation}
    \chi^2 = \left(\vec B-\vec T\right)^{\rm T}\,\tens K^{-1}\,\left(\vec B-\vec T\right)\,,
    \label{eq:chi2}
\end{equation}
where $\vec B$ and $\vec T$ are the theory vectors obtained, respectively, from the benchmark and \CLOE, while $\tens K$ is the covariance matrix constructed following \Euclid\ DR3 specifications and using the benchmark as the fiducial theory predictions.

We impose as a requirement for our comparison that
\begin{equation}
    \chi^2<0.01\,N_{\rm dof}\,,
\end{equation}
which corresponds to requiring the difference between the two codes to be within 10\% of the observational error.

\paragraph{Photometric survey.} Denoting the benchmark spectra as $B_{ij}^{AB}(\ell)$, we compute the observational noise expected for \Euclid\ DR3 as \citepalias{EP-ClOE3}
\begin{align}
    N_{ij}^{\rm GG}(\ell) &= \frac{1}{\bar{n}_i}\delta_{ij}^{\rm K}\,,\\
    N_{ij}^{\rm LL}(\ell) &= \frac{\sigma_\epsilon^2}{2\,\bar{n}_i}\delta_{ij}^{\rm K}\,,
\end{align}
where $\bar{n}_i$ is the number of observed sources in the $i$-th redshift bin and $\sigma_\epsilon$ is the intrinsic ellipticity dispersion.

Following the approach of \citet{EP-Sciotti}, we use $B_{ij}^{AB}(\ell)$ and $N_{ij}^{AB}(\ell)$ to compute the covariance matrix for each of the cases under examination. The diagonal of the covariance is used to obtain the errors $\sigma_{O_i}$ used in \cref{eq:errdist}.

In this work, we use only a Gaussian covariance, neglecting the super-sample and connected non-Gaussian terms, as well as the mode coupling introduced by the mask. For such a reason, we effectively neglect the correlations between different multipoles, obtaining $N_\ell$ matrices $\tens C(\ell)$, independent of each other.

For each multipole, we construct a data vector $D_{ij}^{AB}(\ell)$ extracting a random sample from a multivariate normal with mean $B_{ij}^{AB}(\ell)$ and covariance $\tens C(\ell)$. With this simulated dataset in hand, we can assess the compatibility of the theoretical predictions obtained with \texttt{CLOE}, denoted as $T_{ij}^{AB}(\ell)$, by computing $\chi^2$. This procedure is repeated, generating $N_R=100$ different realisations of the data vector $D_{ij}^{AB}(\ell)$, and we compute the mean value of $\chi^2$. We then assess whether the prediction of \texttt{CLOE} is compatible with the benchmark spectra by comparing the reduced $\chi^2$ with the limiting values corresponding to the chosen probability threshold.

\paragraph{Spectroscopic survey.} Similarly to the previous case, the covariance matrix for the galaxy power spectrum multipoles is derived only considering the Gaussian limit, as described in \cite{Grieb2016}. In this way, the bin-averaged covariance can be written as
\be
    C_{\ell_1\ell_2}(k_i,k_j) = \frac{2\,(2\pi^4)}{V_{k_i}^{\,2}} \,\delta^{\rm K}_{ij} \int_{k_i-\Delta k/2}^{k_i+\Delta k/2} \sigma^2_{\ell_1 \ell_2}(k) \,k^2\, \diff\,k\,,
\ee
where the per-mode covariance is defined as
\be
    \begin{split}
        \sigma^2_{\ell_1 \ell_2}(k)=&\frac{(2\,\ell_1+1)\,(2\,\ell_2+1)}{V_{\rm s}}\\
        &\hspace{10pt}\times \int_{-1}^{+1}\left[P_{\rm gg}(k,\mu)+\frac{1}{\bar{n}}\right]^{\,2} {\cal L}_{\ell_1}(\mu) \,{\cal L}_{\ell_2}(\mu) \,\diff \mu\,,\\
    \end{split}
\ee
as in the corresponding equations of \citetalias{EP-CLOE1} (see Table \ref{tab:func_eqs}). Here, $\mu$ is the cosine of the angle between the wavemode ${\bm k}$ and the line of sight, $\bar{n}$ is the mean number density of the individual spectroscopic bins, $V_{k_i}=4\,\pi\,[(k_i+\Delta k/2)^3-(k_i-\Delta k/2)^3]/3$ is the volume of the spherical shell centred at $k_i$ with width $\Delta k$, and $V_{\rm s}$ is the volume of the considered redshift bin. In this limit, the cross-covariance terms are zero by construction, except for the covariance of different multipoles $P_{\ell}(k)$ at the same wavenumber $k$.

For the purpose of validation, we consider a volume for each spectroscopic bin corresponding to the one already introduced in Sect. \ref{sec:gcspec_bench}, with centres at $z\in\{1,\,1.2,\,1.4,\,1.65\}$ and $\Delta z\in\{0.2,\,0.2,\,0.2,\,0.3\}$. The chosen number density corresponds to the one predicted from a luminosity function of H$\alpha$ galaxies corresponding to Model 3 in \citet{PozHirGea2016}, which was already adopted in other preparation analyses of \Euclid \citep[e.g.][]{EP-Pezzotta}. In this case, we also use the diagonal of the covariance to obtain the errors $\sigma_{O_i}$ used in \cref{eq:errdist}.

A more complete theoretical covariance, also accounting for the impact of the super-survey modes and the trispectrum contribution \citep{WadSco2020} within the geometry defined by the \Euclid angular footprint, will be considered for the analysis of DR1 data. The technical implementation of such covariance will be described in dedicated papers (Euclid Collaboration: Salvalaggio et al. in prep., Euclid Collaboration: Sciotti et al. in prep.). The calculation of the $\chi^2$ statistics is then consistent with that of \cref{eq:chi2}, with the corresponding change of theory and data vectors and covariance matrices.

\section{Validation results}\label{sec:validation}

In this section, we validate the cosmological quantities computed by \CLOE, as well as the observables for the photometric and spectroscopic surveys
across the different configuration cases described in Sect.~\ref{sec:settings}, following the methodology outlined in Sect.~\ref{sec:methods}. For clarity, we divide the validation into separate subsections, focusing individually on cosmological background quantities, photometric observables, and spectroscopic observables.

\subsection{Cosmological functions}\label{sec:cosmology}

Before we focus our attention on the observables computed by \CLOE and their agreement with the external benchmarks, we first want to scrutinise the agreement in the cosmological functions that are the building blocks for such quantities, namely the comoving distance $r(z)$, the normalised Hubble parameter $E(z)=H(z)/H_0$, with $H_0$ the Hubble constant, that is, the present-time value of the Hubble parameter,
the growth factor $g_+(z)$, and the two transverse comoving distances $f_K(z)$ and $f_K(z_1,z_2)$, accounting for possible spatial curvature. As shown in Sect.\ 2 of 
\citetalias{EP-CLOE1}, 
these quantities enter all the predictions made by \CLOE, and therefore any difference would propagate to the final observables.

\begin{figure}[h]
    \centering
    \includegraphics[width = 0.95\columnwidth]{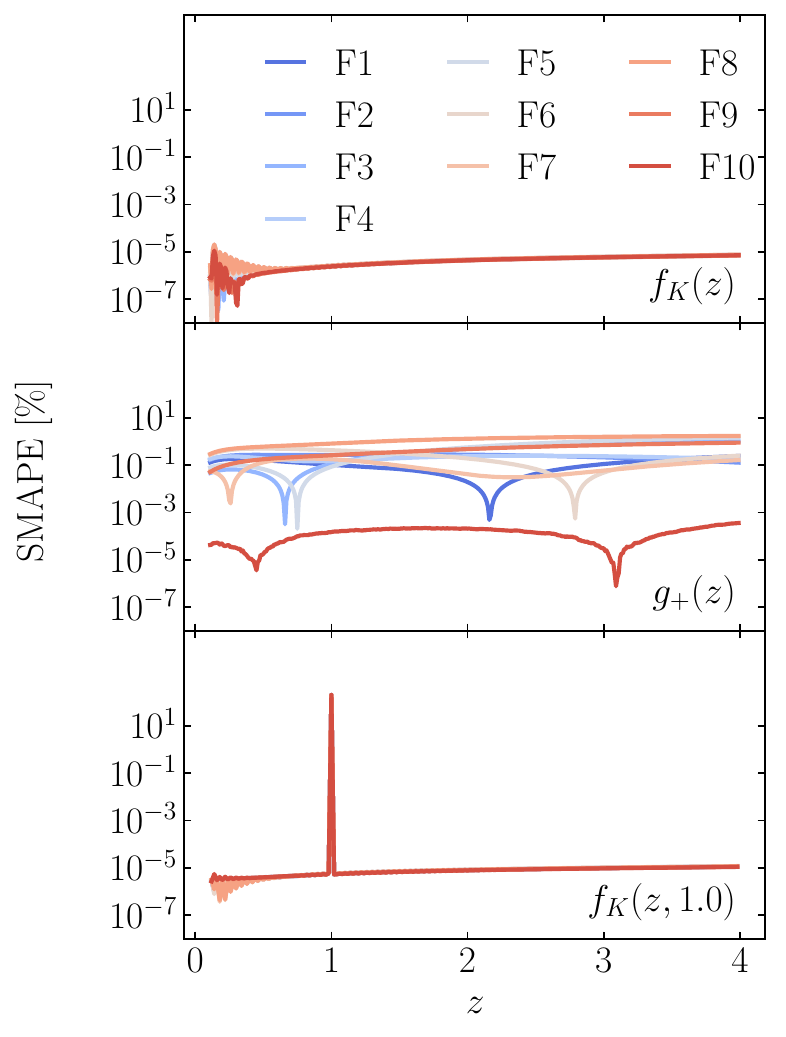}
    \caption{Comparison of the cosmological quantities internally computed by \CLOE with those extracted from \camb for the cosmologies listed in \cref{tab:cosmopars}. Please note that \camb automatically sets $f_K(z_1,z_2)=0$ when $z_1>z_2$, while \CLOE returns its absolute value for any redshift combination. In order to perform this comparison, we symmetrise the results obtained from \camb. The spikes visible in the comparison of the $f_K(z_1,z_2)$ values are due to this function vanishing when $z_1=z_2$. We only show the transverse distance $f_K(z_1,z_2)$ with $z_2=1$, with the other cases showing the same behaviour.}
    \label{fig:CAMB_comparison}
\end{figure}

\CLOE extracts $E(z)$ and $r(z)$ directly from a Boltzmann solver (either \camb or \class), while it computes the other quantities internally. Given that both \camb and \class have been extensively validated, we can safely assume that the quantities that we retrieve from these codes do not present any issue. However, it is necessary to assess the reliability of the cosmological quantities computed by \CLOE, namely $g_+(z)$ and the comoving distances $f_K(z)$ and $f_K(z_1,z_2)$. We present this comparison in \cref{fig:CAMB_comparison}, where we show, for all cosmologies considered, the SMAPE of the cosmological quantities internally computed by \CLOE with respect to those which can be retrieved from \camb.\footnote{We refer the reader to \citet{Torrado:2020dgo} for a comparison between \camb and the quantities which can be extracted by interfacing \cobaya with Boltzmann solvers.}

We notice that the distance functions $f_K(z)$ and $f_K(z_1,z_2)$ agree very well with those computed by \camb, with the SMAPE always staying several orders of magnitude below the $1\%$ level. Furthermore, the change in cosmological parameters does not lead to any significant difference in the comparison, as can be seen by the overlap of the lines corresponding to different cosmologies in \cref{fig:CAMB_comparison}. However, we observe larger differences in the growth factor $g_+(z)$. This is because while \camb obtains this quantity by solving the differential equation for matter perturbation $\delta_{\rm m}$, \CLOE computes $g_+(z)$ by taking the ratio of the matter power spectra at different redshifts and at a reference scale 
\citepalias[see Eq.\ 22 of][for details on this calculations]{EP-CLOE1}. 
Despite these differences, the comparison still yields results within  $1\%$ accuracy. However, a revised version of \CLOE will aim to further improve this comparison by extracting this quantity directly from the Boltzmann solvers through \cobaya.

It is important to stress that throughout this validation we use \CLOE interfaced with \camb. As discussed in 
\citetalias{Paper2}, 
\CLOE can also be interfaced with \class. However, at the time our comparison was performed, our baseline recipe for non-linear corrections, \texttt{HMCode2020} \citep{Mead:2020vgs}, was not available in \class, making it impossible to apply the full validation pipeline by interfacing \CLOE with \class. Nevertheless, in Appendix\ \ref{sec:class_comp} we show a comparison between the main cosmological quantities used in the calculation of theoretical predictions between \CLOE+~\class and \CLOE+~\camb.

\subsection{Photometric 3\texttimes2pt observables}\label{sec:photo_validation}

In this section, we apply the validation metrics discussed in Sect. \ref{sec:methods} to the photometric observables, namely the $\CABij(\ell)$ power spectra for all tracers and redshift bin combinations \citepalias[see Sect.\ 3 of][]{EP-CLOE1}. We refer the reader to Appendix \ref{app:photo_int} for a discussion of the intermediate quantities entering the calculation of the $\CABij(\ell)$.

\begin{figure*}[th!]
    \centering
    \begin{tabular}{cc}
        \includegraphics[width = 0.95\columnwidth]{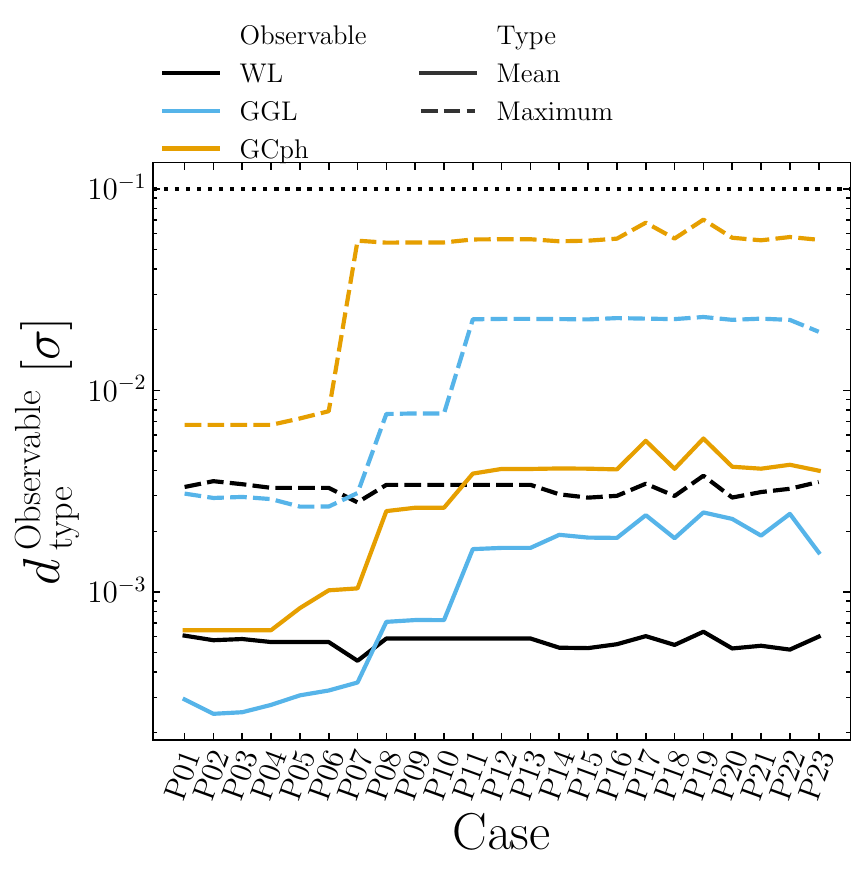} &  
        \includegraphics[width = 0.95\columnwidth]{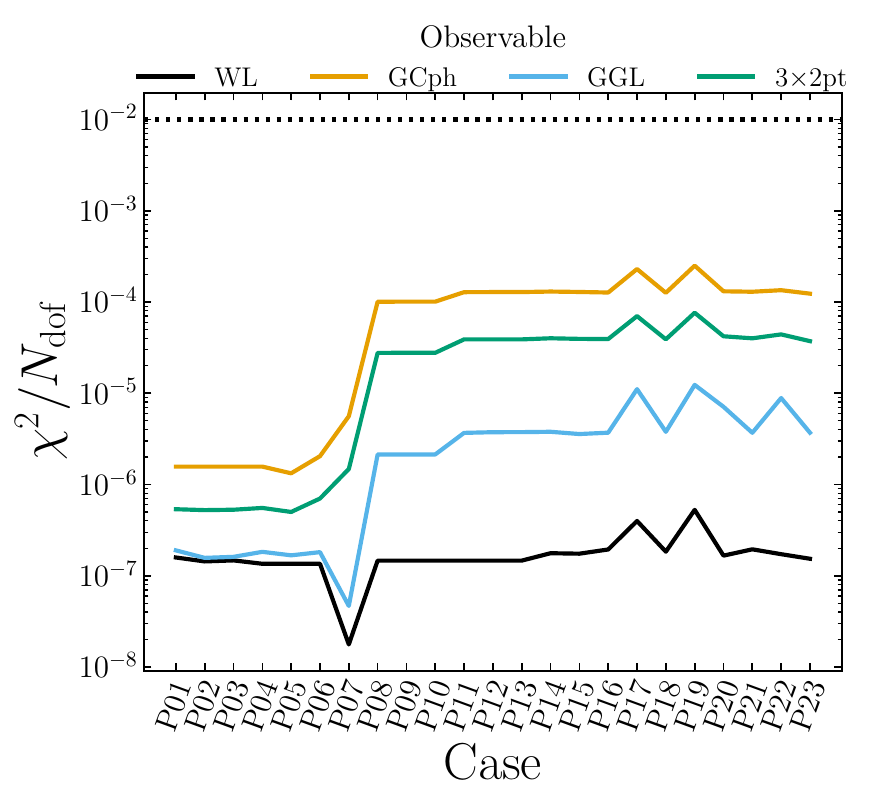}\\ 
    \end{tabular}
    
    \caption{\textit{Left panel:} Distance in units of the expected error for the three photometric observables, both when averaged over the multipoles (solid lines) and taking the highest value of the distance (dashed lines). The dotted black line indicates our threshold of $d=0.1\,\sigma$. \textit{Right panel:} Solid lines show the reduced $\chi^2$ value for WL (black), GCph (orange), GGL (cyan), and their combination (green). The dashed line represents the limiting value of the reduced $\chi^2$ that we choose to be 0.01. The results are shown for the different cases described in \cref{tab:cases}.}
    \label{fig:obs_summary}
\end{figure*}

In \cref{fig:obs_summary}, we show in the left panel the mean and maximum distances for each of the validation cases, while in the right panel we report the value of the reduced $\chi^2$ (solid lines), obtained by dividing the results of \cref{eq:chi2} by the number of degrees of freedom $N_{\rm dof}$ corresponding to each case. We show the trend of such a quantity for the two photometric observables (WL and photometric galaxy clustering, GCph), their cross-correlation (galaxy-galaxy lensing, GGL) and, for the $\chi^2$ calculation, their combination (3$\times$2pt). We find that, in all validation cases, both the distance in units of the expected error and the reduced $\chi^2$ fall within the threshold values, highlighting the statistical compatibility of the theoretical predictions of \texttt{CLOE} with those of the benchmark code. Furthermore, we can see how the two tests yield very similar results, highlighting how the $\chi^2$ test reaches, despite accounting for the off-diagonal terms of the covariance, the same performance as the distance test, where such terms are neglected.

We notice how all observables exhibit better agreement for the P01--P06 cases. This is because these cases assume the galaxy distribution of ISTF, which is a distribution that is split into fewer bins and whose functions are much smoother than in other validation cases. This allows us to reduce numerical and interpolation errors with respect to other galaxy distributions when computing the kernel functions, thus yielding a better comparison. We refer the reader to Appendix \ref{app:photo_int} for a visualisation of kernel validation in the different cases. Overall, we find no significant discrepancy between the benchmark and the predictions of \texttt{CLOE}, independently of the observable analysed or of the validation case.

\subsection{Spectroscopic galaxy clustering observable}\label{sec:spectro_obs}

In this section, we validate the implementation of the model relevant for the spectroscopic probe of \Euclid. The validation of intermediate quantities, such as the individual contributions entering the perturbative expansion of the reference non-linear model implemented in \CLOE, will be presented in a dedicated paper (Euclid Collaboration: Crocce et al. in prep.). Here, we restrict our analysis to the final spectroscopic observables, namely the Legendre multipoles $P_\ell(k,z)$ of the anisotropic galaxy power spectrum.

\begin{figure}
    \centering
    \includegraphics[width=\columnwidth]{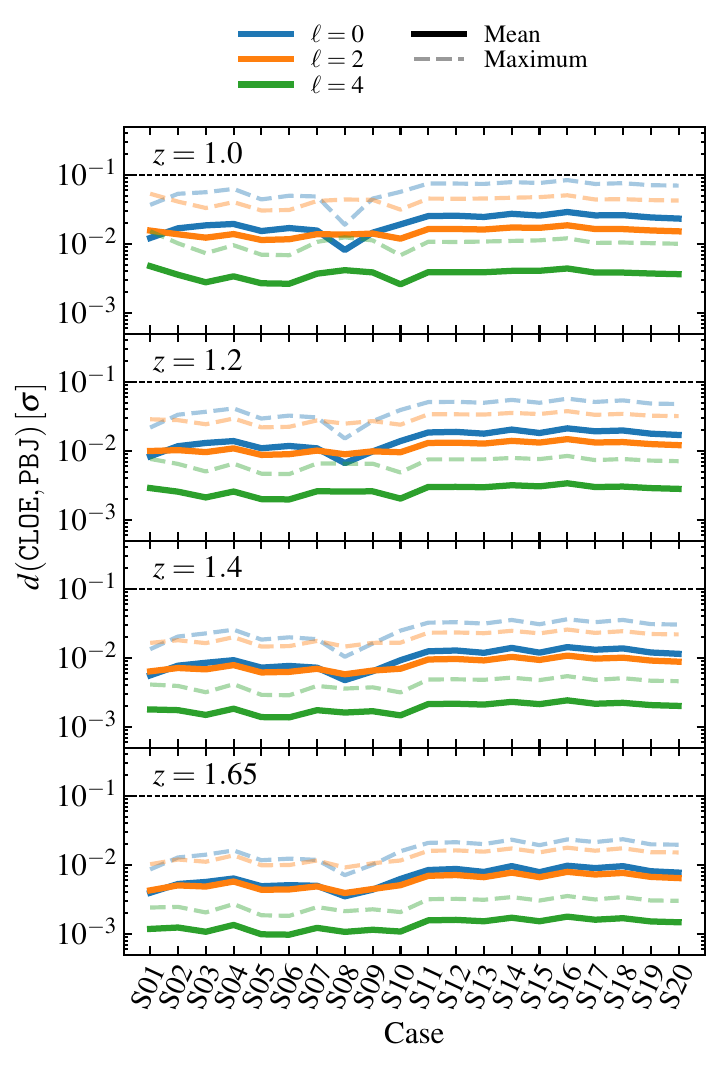}
    \caption{Maximum relative difference between \CLOE and \pbj in units of a Gaussian standard deviation obtained with the configuration described in Sect. \ref{sec:err_comparison}. Different panels correspond to different redshifts, as shown in the corresponding top left corners. Different multipoles are identified by the different colours, while solid and dashed lines mark the average and maximum difference along the sampled $k$ values, respectively. The dotted horizontal line in each panel marks the threshold of $10\%$ of the statistical error.}
    \label{fig:summary_spectro}
\end{figure}

\begin{figure}
    \centering
    \includegraphics[width=\columnwidth]{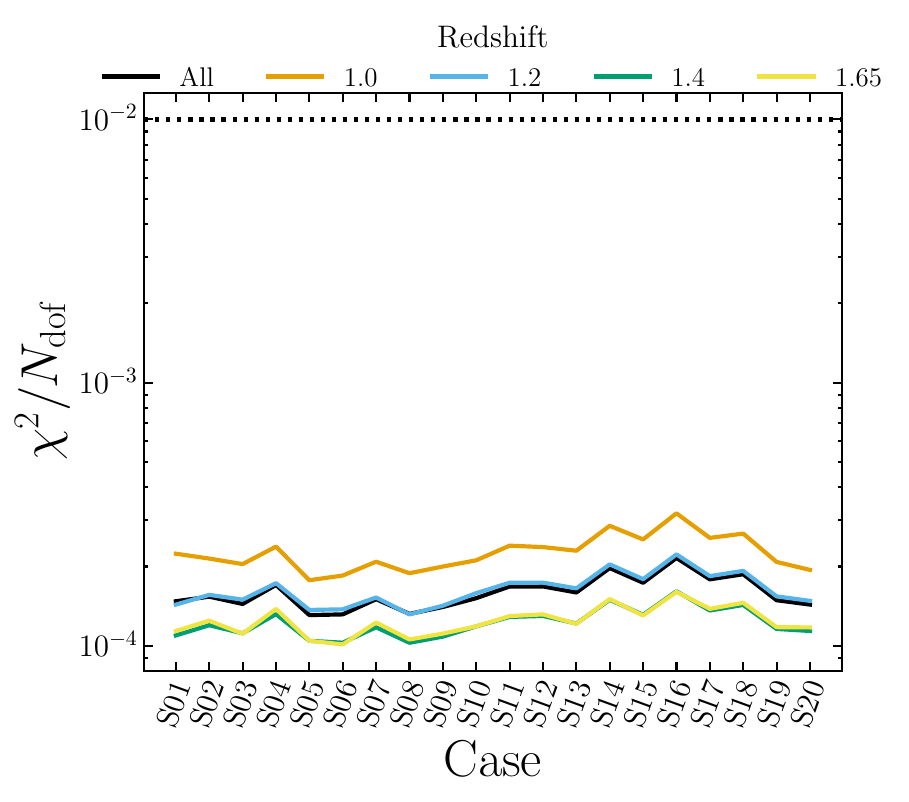}
    \caption{Reduced $\chi^2$ displayed in the 20 validation cases for the GCsp probe combination. As in \cref{fig:obs_summary}, the dotted line shows the limiting threshold of the $\chi^2$, corresponding to the requirement of \CLOE\ departing from the benchmark by less than 10\% of the observational error.}
    \label{fig:summary_spectro_chi2}
\end{figure}

The main results of the validation are presented in \cref{fig:summary_spectro}. Here we show the distance metrics between \CLOE and \pbj as defined in \cref{eq:mean_errdist,eq:max_errdist}, considering all redshift bins, multipole orders, and validation cases. Since the recipe for the anisotropic galaxy power spectrum $\Pgg(k,\mu)$ in \CLOE follows the same implementation as in \pbj, we find a good match between the two sets of predictions, with a relative error that always stays below 10\% of the statistical error. This happens not only when considering the mean distance between the two sets of predictions but also with the maximum distance. Given this level of agreement, the average reduced $\chi^2$ values obtained by comparing the theory vectors generated by \CLOE and \pbj are also extremely good, as we show in \cref{fig:summary_spectro_chi2}. Quantitatively, the average $\chi^2$ is entirely within the $68\%$ confidence region of the corresponding $\chi^2$ distribution with the same number of degrees of freedom. Residual fluctuations are primarily due to differences in the unit conventions adopted by the two codes (i.e. choosing $\mathrm{Mpc}^{-1}$ against $h\,\mathrm{Mpc}^{-1}$), which can induce small discrepancies in the computation of the internal components of the loop expansion, as well as in the algorithms used to project the 2D power spectrum onto the Legendre polynomials. While this section focuses only on the distance metrics and averaged $\chi^2$ statistics, a direct comparison of the theory vectors for a reference case is presented in Appendix~\ref{sec:app_spectro}.

\section{Comparison of real-space correlation functions}\label{sec:corrfuncs}

In Sect.\ \ref{sec:photo_validation} and \ref{sec:spectro_obs}, we assessed the validity of \CLOE focusing on the harmonic-space power spectra $\CABij(\ell)$ and the Legendre multipole $P_{\mathrm{obs}, \ell}(k^{\mathrm{fid}}, z)$ as the observables to be compared. However, several analyses of LSS data in the literature rely instead on two-point correlation functions in real space. For the photometric survey, these are the galaxy correlation function $\xiGGij$, the two lensing correlations $\xiplusij$ and $\ximinusij$, and the galaxy-galaxy lensing correlation $\xiGLij$, while for the spectroscopic survey, we are interested in the 2PCF Legendre multipoles $\xi_{{\rm obs},\ell}(s^{\rm fid},z)$.

\begin{figure*}[h]
    \centering
    \includegraphics[width = 0.95\textwidth]{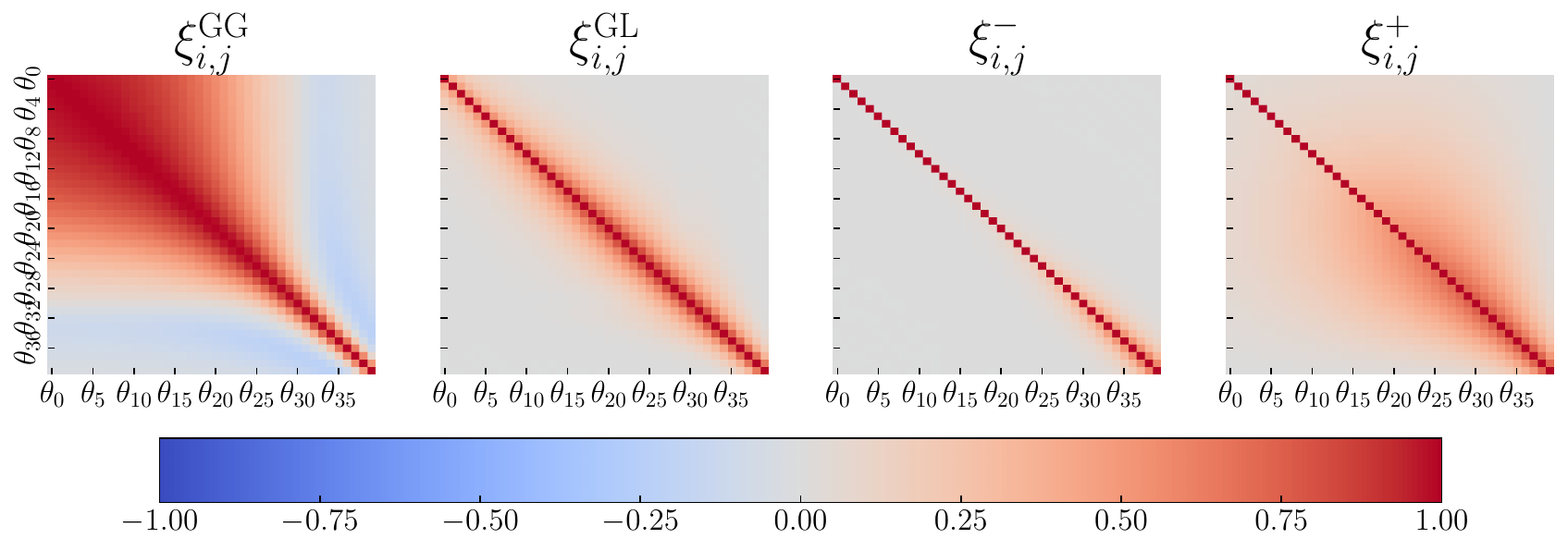}
    \caption{Correlation matrices between the different angles $\theta_i$ for which the 2PCF are computed, taken for $i=j=6$ in the P23 case.
    }
    \label{fig:2PTCF_photocov}
\end{figure*}

\begin{figure}
    \centering
    \includegraphics[width=\columnwidth]{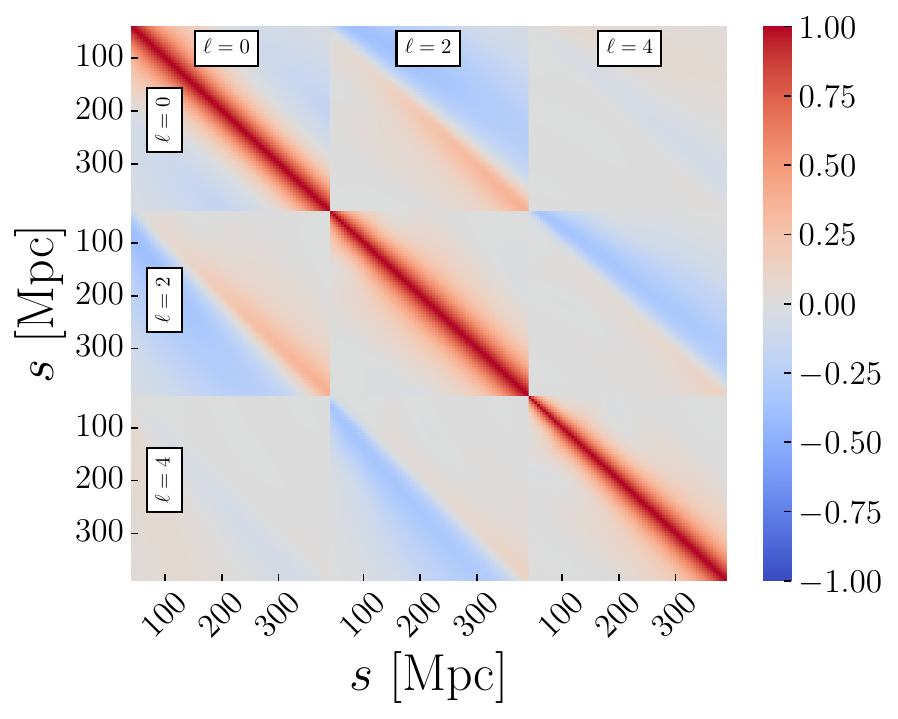}
    \caption{Correlation matrix of the linear-theory 2PCF multipoles assuming the setup adopted for the validation, thus linear bias and RSD signal, and number density corresponding to the ones defined in 
    \citetalias{EP-ClOE3}.}
    \label{fig:2PCF_correlation}
\end{figure}

In the comparison performed in Sect.\ \ref{sec:validation}, we relied on the intuitive definition of distance defined in \cref{eq:errdist}, obtaining the error $\sigma_{O_i}$ from the diagonal of the covariance matrices. However, this requires that the off-diagonal terms in the covariances are negligible, which is not the case for 2PCF \citep[see e.g.][]{Eisenstein2001}. This can be easily observed in \cref{fig:2PTCF_photocov,fig:2PCF_correlation}, where we show the correlation matrix for the photometric and spectroscopic cases.

This implies that the error obtained by taking the diagonal of these matrices would yield a significant underestimation of the uncertainties with respect to what would be obtained by considering the full covariances. For such a reason, in comparing the photometric and spectroscopic 2PCF we rely only on the $\chi^2$ estimate of \cref{eq:chi2}, avoiding the computation of the distance in units of error.

Furthermore, \texttt{CLOE} computes the 2PCF by projecting the observables validated in Sect.\ \ref{sec:validation}. For such a reason, we do not repeat the validation for all the cases discussed in Sect.\ \ref{sec:settings}, deeming it sufficient to perform the validation in the reference cases, P23 and S01 for the photometric and spectroscopic surveys, respectively.

For the photometric correlation functions, we perform the validation by computing the covariance matrix using the public \texttt{OneCovariance} code \citep{Reischke:2024fvk}.\footnote{\url{https://github.com/rreischke/OneCovariance}} This computes the covariance in harmonic space and then projects it into real space using the equations presented in \citet{Joachimi2021}. 

For the spectroscopic correlation functions, similar to the case in Fourier space, we only make use of the Gaussian limit, which translates to a conservative validation of the implementation in \CLOE. In this case, the covariance matrices for the four different spectroscopic bins have been computed using the public \texttt{GaussianCovariance} code.\footnote{\url{https://gitlab.com/veropalumbo.alfonso/gaussiancovariance}}

We present the results of our comparison in \cref{tab:chi2_xi}, where it is possible to notice how the reduced $\chi^2$ value obtained for 2PCF is within the expected limits for both photometric and spectroscopic surveys, with the exception of $\xiGGij$, for which the difference between \CLOE\ and the benchmark is within 20\% of the observational error, rather than the required 10\%. While such a discrepancy will not affect constraints in a significant way, in future developments we will assess the possible numerical differences that yield such a result.

\begin{table}
\centering
\caption{Reduced $\chi^2$ for 2PCF of photometric (top row) and spectroscopic (bottom row) observables.}
\begin{tabular}{c||c|c|c|c} 
\hline
&\multicolumn{4}{c}{Photometric}\\
\hline
& $\xiGGij$ & $\xiGLij$ & $\ximinusij$ & $\xiplusij$ \\
 \hline 
$\chi^2/N_{\rm dof}$ & 0.040 & 0.001 & $6\times10^{-5}$ & $5\times10^{-5}$ \\
\hline
\hline
&\multicolumn{4}{c}{Spectroscopic}\\
\hline
& $z=1.0$ & $z=1.2$ & $z=1.4$ & $z=1.65$ \\
 \hline 
$\chi^2/N_{\rm dof}$ & $7\times10^{-4}$ & $5\times10^{-4}$ & $4\times10^{-4}$ & $4\times10^{-4}$ \\
\hline
\end{tabular} 
\label{tab:chi2_xi}
\end{table}

\section{Conclusions}\label{sec:conclusions}

The main objective of this work was to benchmark the code used to obtain theoretical predictions for \Euclid\ observables: \CLOE. In order to do so, we compared these predictions with those obtained from external software that implements the same recipe.

Although we do not explore the parameter space systematically, we performed this benchmarking in a wide set of validation cases, described in Sect. \ref{sec:settings}. This allowed us to test the features implemented in \CLOE, which are described in detail in \citetalias{EP-CLOE1}.

After presenting the methodology to quantify the discrepancy between \CLOE and the benchmark in Sect.\ \ref{sec:methods}, and the external codes used as benchmarks in Sect.\ \ref{sec:benchcodes}, we compared as a first step the basic cosmological ingredients of our recipe in Sect.\ \ref{sec:cosmology}. Here we found a sub-percent agreement between \CLOE and \life for all functions in all the chosen cosmologies and, even more importantly, we found a very good agreement between \CLOE and \camb. For the latter comparison, we found that the most significant difference is obtained for the growth factor $g_+(z)$; the internal calculation of \CLOE approximates this with the square root of a ratio of matter power spectra, which, when compared with the \camb prediction, yields a discrepancy of 1 percent. Following the validation of these fundamental ingredients, we focused separately on the two surveys of \Euclid: spectroscopic and photometric.

In Sect.\ \ref{sec:spectro_obs}, we discussed the validation of the final spectroscopic observables against an external code, \pbj, which implements the same model for the spectroscopic galaxy power spectrum based on the EFTofLSS framework. For this comparison, we only considered the final \Euclid observables, the power spectrum Legendre multipoles, while we left the benchmarking of intermediate quantities to a dedicated work (Euclid Collaboration: Crocce et al. in prep., Euclid Collaboration: Moretti et al. in prep.). We computed a distance metric between the two codes based on a \Euclid-like statistical uncertainty that assumes only the Gaussian limit, therefore providing a much more conservative comparison than if also including non-Gaussian contributions. We found an optimal agreement between \CLOE and \pbj, with a distance metric that stays consistently below the 10 percent expected statistical error for all redshifts and wavemodes. Furthermore, the $\chi^2$ analysis shows that the theoretical predictions of \CLOE are in statistical agreement with the dataset generated using \pbj.

The validation of the photometric observables was presented in Sect.\ \ref{sec:photo_validation}, where we obtained the results of the distance and $\chi^2$ tests for the calculation of  $\CABij(\ell)$, comparing the calculations of \CLOE\ with those of \life. We show in \cref{fig:obs_summary} how, in all our validation cases, both the distances and the reduced $\chi^2$ are well within their limiting values for the chosen probability threshold ($68\%$). We find no significant differences for the separate photometric observables or their combination, as well as no significant trend with the analysed cases, except for a better agreement found in the case P01--P06, where we use a smooth galaxy distribution. Thus, we can conclude that the harmonic power spectra of \CLOE are statistically compatible with the benchmark within the expected DR3 errors of \Euclid.

In addition to the final observables for this survey, we also validated the implementation of intermediate calculations. We presented in Appendix \ref{app:photo_int} the validation of the kernel functions $\WAi(z)$, the power spectra for galaxies, matter, and intrinsic alignment. We found these functions to be compatible with the benchmark well within 1 percent, with the worst performance found for the power spectra containing the intrinsic alignment terms, which are those related to the growth factor $g_+(z)$.

In \cref{sec:reference}, we showed a more detailed comparison for a single case. For the photometric observables, we chose the P23 case as a reference, showing in \cref{fig:distance_reference} the distance in units of the observational errors as a function of the multipole $\ell$, for each observable. Similarly, in \cref{fig:GCspec_dist_S01} we showed the trends in $k$ of the distance for the spectroscopic observables, choosing as a reference the S01 case.

Similarly, the validation of the 2PCF Legendre multipoles points towards an optimal consistency of the implementation present in \CLOE with the one produced by \coffe. 
In this case, given the non-negligible contributions of the off-diagonal terms in the covariance, we only performed the $\chi^2$ analysis, reported in \cref{tab:chi2_xi}, finding good statistical agreement between the \CLOE predictions and the benchmark.

In \cref{tab:chi2_xi}, we also report the results of the $\chi^2$ analysis for photometric 2PCFs, obtained as a projection of  $\CABij(\ell)$. In this case, we also find good agreement with the benchmark: all observables are well within our chosen threshold, with the exception of $\xiGGij$ whose difference with respect to \life is within 20\% of the statistical error, rather than the required 10\%. Given that our requirements are quite stringent, we are confident that such a difference will not propagate to biases on parameter estimation. However, our future goal is to investigate the source of this difference to reduce it to the required threshold.

Thanks to the results obtained in this paper, we conclude that \texttt{CLOE} is reliable and accurate software, able to compute \Euclid's main observables efficiently while in agreement with other software. We are therefore confident that the final results obtained with \Euclid will have negligible bias resulting from the theoretical calculations we examined in this work.

\begin{acknowledgements}
\AckEC 
MM acknowledges funding by the Agenzia Spaziale Italiana (\textsc{asi}) under agreement no. 2018-23-HH.0 and support from INFN/Euclid Sezione di Roma. SC acknowledges support from the Italian Ministry of University and Research (\textsc{mur}), PRIN 2022 `EXSKALIBUR – Euclid-Cross-SKA: Likelihood Inference Building for Universe's Research', Grant No.\ 20222BBYB9, CUP D53D2300252 0006, and from the European Union -- Next Generation EU. SD acknowledges support from the Italian Ministry of University and Research (\textsc{mur}), PRIN 2022 `LaScaLa - Large Scale Lab', Grant No.\ 20222JBEKN, founded by the European Union -- Next Generation EU. During part of this work, AMCLB was supported by a Paris Observatory-PSL University Fellowship, hosted at the Paris Observatory. SP acknowledges support through the \textit{Conception Arenal Programme} of the Universidad de Cantabria and funding from the project UC-LIME
(PID2022-140670NA-I00), financed by MCIN AEI/ 10.13039/501100011033/FEDER, UE. The authors acknowledge the contribution of the Lorentz Center (Leiden), and of the European Space Agency (ESA), where the workshop "Making \CLOE shine" and the "\CLOE workshop 2023" were held.

\end{acknowledgements}

\bibliographystyle{aa}
\bibliography{joint_biblio} 

\begin{appendix}

\section{Reference case comparison}\label{sec:reference}

In Sects. \ref{sec:photo_validation} and \ref{sec:spectro_obs} we compare the predictions of \texttt{CLOE} with those of the benchmark in a variety of cases, exploring different features that are available in \texttt{CLOE}. Here, instead, we present the comparison in more detail in a specific case, that is, the one used as a fiducial to obtain the results presented in \citetalias{EP-ClOE3}.

\subsection{Photometric survey}

\begin{figure}[h]
    \centering
    \includegraphics[width = 0.95\columnwidth]{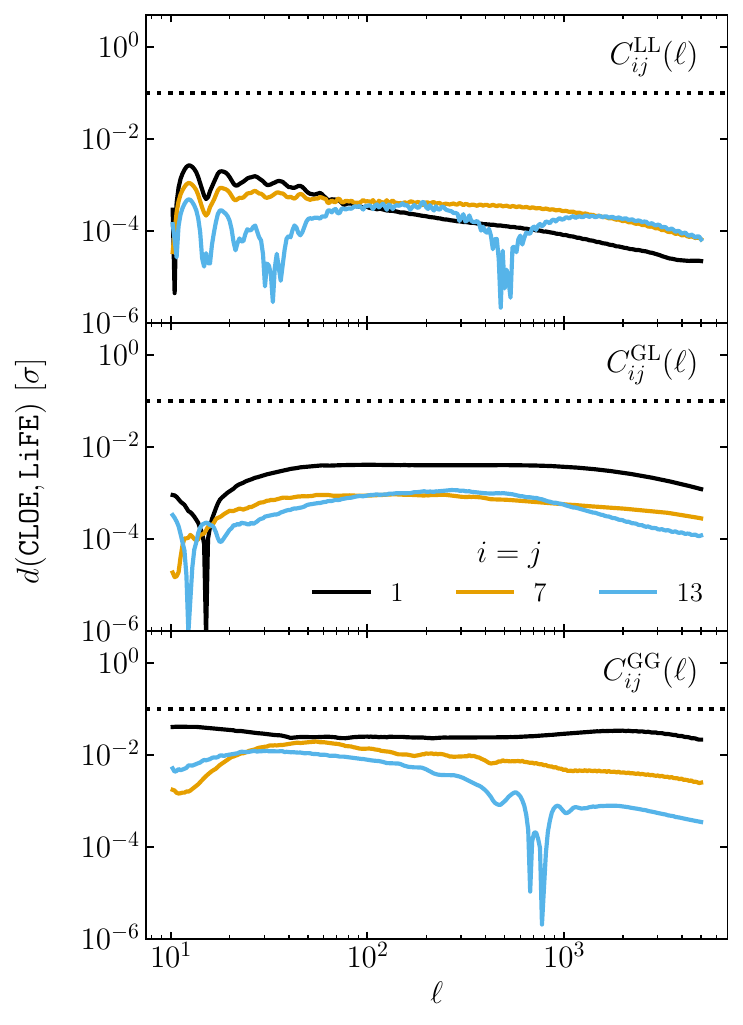}
    \caption{Distance in units of the expected error as a function of the multipole $\ell$ for WL ({\it top panel}), GCph ({\it bottom panel}), and their cross-correlation GGL ({\it middle panel}). The different colours indicate the redshift bin considered. The black dotted line shows the threshold value of $0.1\,\sigma$. This comparison is performed in the reference case P23.}
    \label{fig:distance_reference}
\end{figure}

For the photometric survey, we choose as a reference the P23 case, and we show in \cref{fig:distance_reference} the trends in multipoles of the distance in units of the error for WL, GCph, and their cross-correlation GGL.
We find no significant trend with the multipoles $\ell$ and no significant difference between the analysed probes. We find GCph to be the observable with the highest values of this distance, compatible with the results presented in Sect.\ \ref{sec:photo_validation}, which is a direct result of the smaller observational error for this observable with respect to the others.

\subsection{Spectroscopic survey}
\label{sec:app_spectro}

\begin{figure}[t]
    \centering
    \includegraphics[width=\columnwidth]{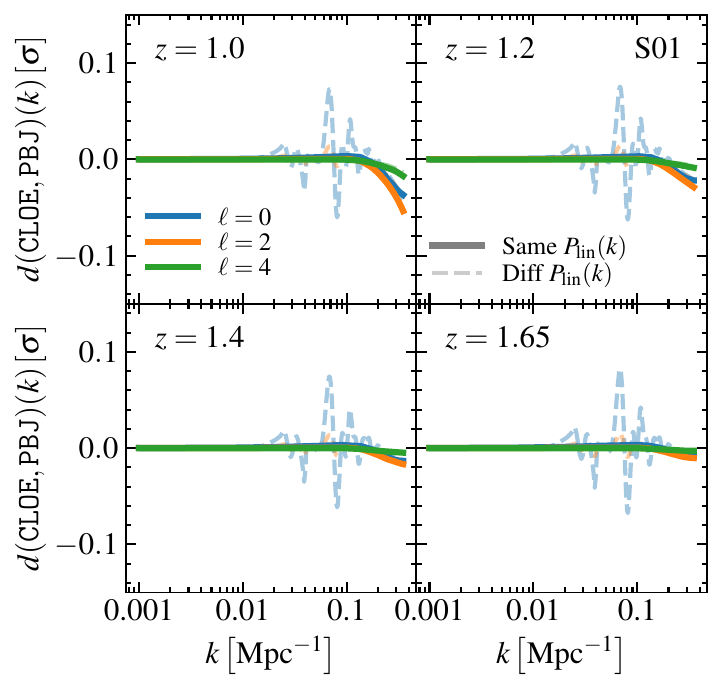}
    \caption{Systematic difference between \CLOE and \pbj for the case S01 (flat \LCDM) of GCsp. The difference is quantified in terms of the distance metric defined in Eqs.~(\ref{eq:max_errdist}) and (\ref{eq:mean_errdist}). Different colours correspond to different multipoles, while solid and dashed lines denote the case when \CLOE and \pbj use the same input linear power spectrum or not, respectively.
    }
    \label{fig:GCspec_dist_S01}
\end{figure}

In \cref{fig:GCspec_dist_S01}, we show the distance metric as a function of the wavemode $k$ for a reference case, a flat \LCDM cosmology as in case S01. In all panels, we show the accuracy between \CLOE and \pbj using two sets of predictions: in the first one, each code obtains the linear power spectrum predictions from an individual call to the Boltzmann solver (\camb in this case), while, in the second one, the same linear predictions are used consistently across the two codes. In the first case, we observe residuals in the final Legendre multipoles that are always smaller than 10\% of the corresponding statistical error for a few spurious positions along the $k$ axis, mostly concentrated over the BAO scales. This is a consequence of employing slightly different versions and accuracy flags in the Boltzmann solver, but despite this difference, the trend is to have an optimal consistency in terms of the broadband of the multipoles. The agreement is even better when considering the same input spectra, for which we only observe a minor deviation at $k\gtrsim0.1\kMpc$ that is induced by a small difference in the way infrared resummation is carried out in the two codes. This is mostly due to the different set of units adopted by the two codes, with or without $h$, respectively (Euclid Collaboration: Moretti et al. in prep.).

\section{Photometric intermediate quantities}\label{app:photo_int}

In Sect. \ref{sec:photo_validation}, we quantified the departures of CLOE predictions for photometric observables with respect to the chosen benchmark code. Here, instead, we apply our validation methodology to the intermediate quantities related to the photometric survey of \Euclid, mainly the kernel functions and the power spectra 
\citepalias[see Sect.\ 3 of][]{EP-CLOE1}. In this section, we show the validation results across the cases described in Sect. \ref{sec:settings}, highlighting the overall comparison of these quantities.

The first set of intermediate quantities that we focus on are the kernel functions entering the expression for photometric observables, namely the shear kernel $\Wgammai(z)$, the magnification kernel $\Wmui(z)$, the galaxy kernel $\Wgali(z)$, and the intrinsic alignment kernel $\WIAi(z)$.

\begin{figure}[h!]
    \centering
    \includegraphics[width = 1.\columnwidth]{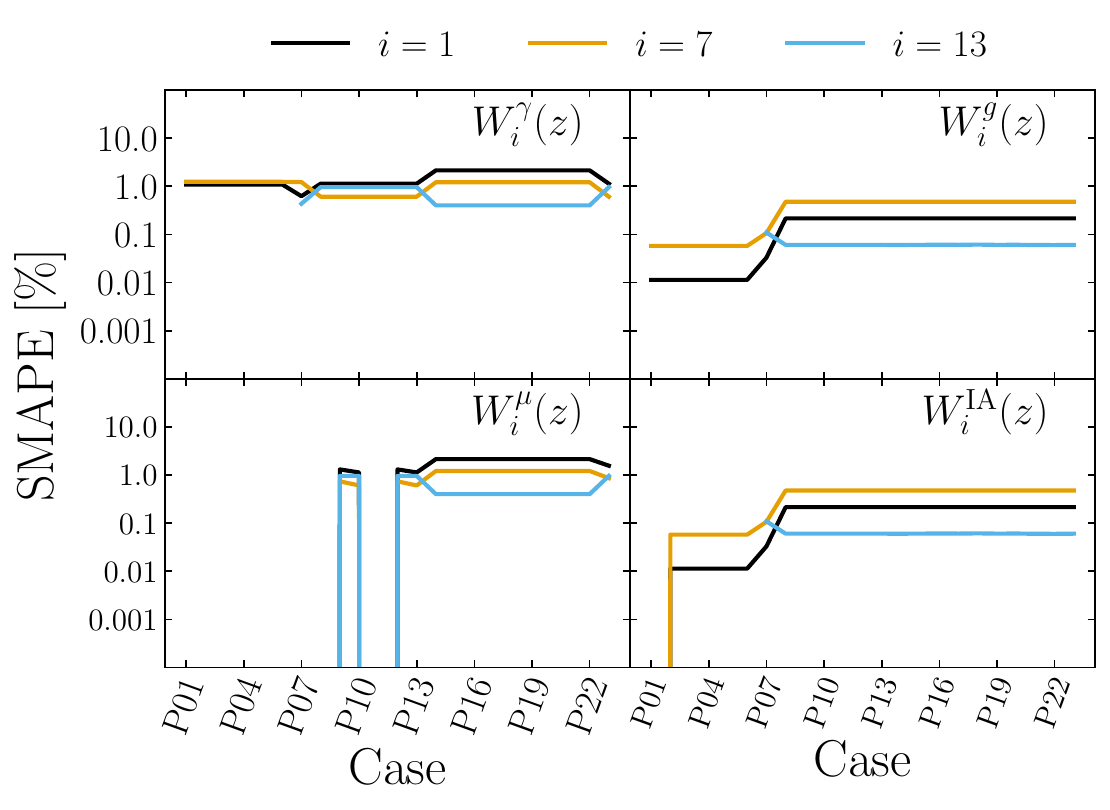}
    \caption{Summary SMAPE values for the kernels comparison between \CLOE and \life in the different cases investigated. Each colour refers to a different redshift bin of the tomographic \Euclid analysis. The figure shows the comparison result in each case for representative redshift bins at low ($i=1$, black lines), intermediate ($i=7$, orange lines) and high redshift ($i=13$, cyan lines). Notice that for the cases P01 to P06, this latter bin is not available, as the $n(z)$ is divided into only 10 tomographic bins. The specifications of the different cases are reported in \cref{tab:cases}.}
    \label{fig:kernel_summary}
\end{figure}

We report the overall value of the SMAPE for the four kernels in \cref{fig:kernel_summary}. Here we can notice how the SMAPE values are of the order of one per cent, with $\Wgammai$ and $\Wmui$ being the kernels with the worst agreement. This deviation is caused by the fact that the calculation of these functions requires an integration of the binned galaxy redshift distribution
\citepalias[see][]{EP-CLOE1};
if such a distribution is not smooth, differences in the numerical sampling between the codes can propagate to the final kernel computations. Indeed, it can be noticed how the cases P01--P06 yield a generally better comparison, as these cases assume the ISTF galaxy redshift distribution \citep{Blanchard-EP7}, which is more regular and with broader bins with respect to the other two, limiting the impact of numerical errors.
    
Following the validation of the kernel functions, we shift our attention to the power spectra entering the calculation of the photometric observables 
\citepalias[see section 3.1 of][]{EP-CLOE1}.
These are the matter power spectrum $\Pmm(k,z)$, the galaxy power spectrum $\Pgg(k,z)$, and the intrinsic alignment power spectrum $\Pii(k,z)$, together with all cross combinations $\Pgm$, $\Pgi$, and $\Pmi$.

\begin{figure}[h!]
    \centering
    \includegraphics[width = 1.\columnwidth]{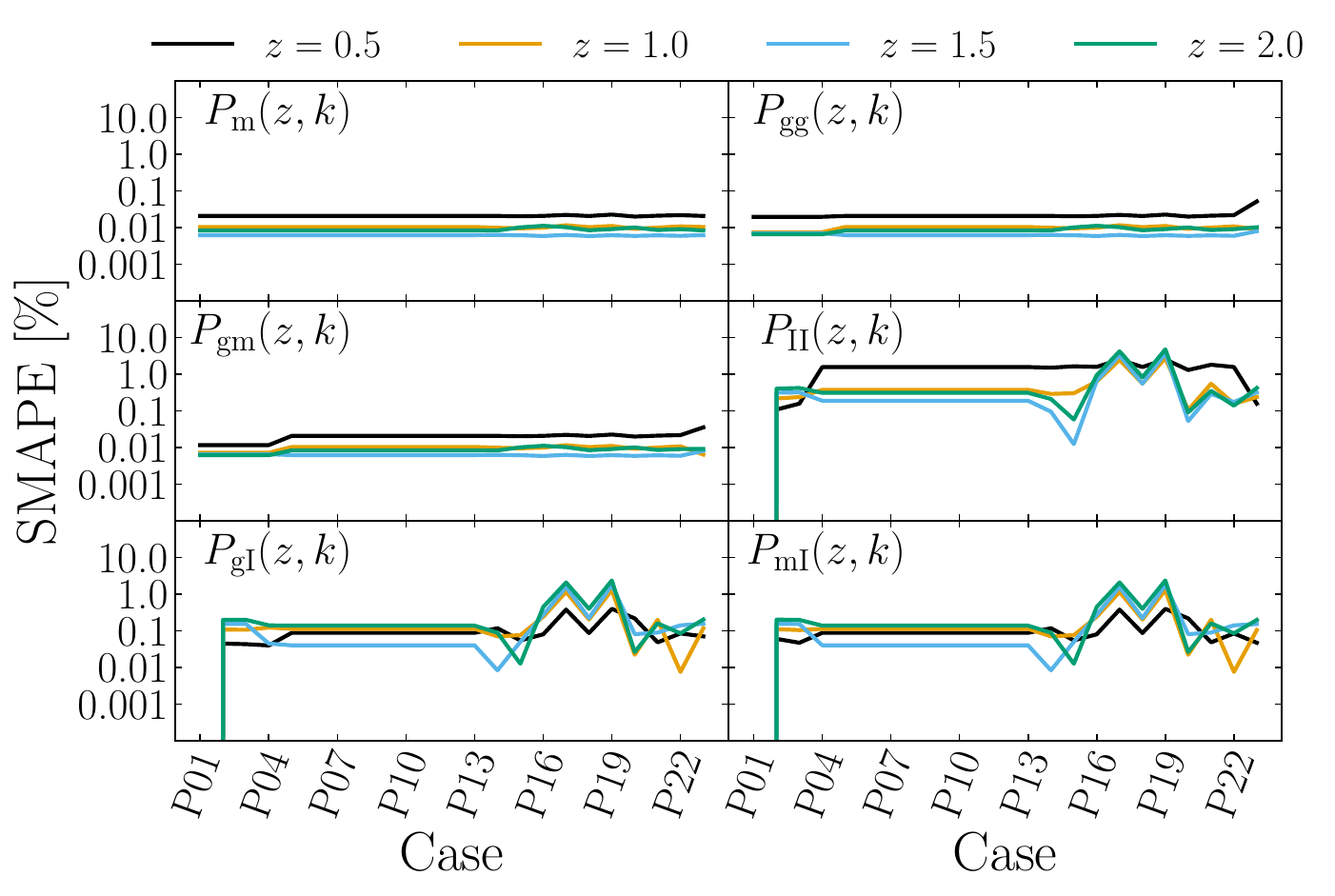}
    \caption{Summary comparison for the intermediate power spectra, as computed by \CLOE and \life. For each of the different power spectra, shown in separate panels, we show the comparison at different redshifts. The specifications of the different cases are reported in \cref{tab:cases}.}
    \label{fig:power_spectra_summary}
\end{figure}

We report the overall value of the SMAPE for the six power spectra, computed at four different redshifts, in \cref{fig:power_spectra_summary}. It is possible to notice how the comparison does not exhibit a strong case dependency, with mostly constant trends for the SMAPE values at all redshifts and for all spectra. Nevertheless, it is possible to see a dependency on the chosen case for the P13--P22 cases, where the cosmological parameters change in value, in the power spectra containing the intrinsic alignment contribution. This trend can be connected to the differences between the two codes in the growth factor $g_+(z)$, discussed in Sect.\ \ref{sec:cosmology}.

\section{Comparison with \ccl}\label{sec:pyccl_comp}

The comparison against \ccl has proceeded as described above for \life. To compute the final observables of interest (2PCF and power spectra), the code requires as external input the $n(z)$, the tabulated values of galaxy and magnification bias, and the tabulated IA kernel. The $\CABij(\ell)$ are then computed, using the Limber approximation, using quadrature integration (which is the most accurate and for which the best agreement is found) as mentioned in Sect.\ \ref{sec:CCL}. 
To be as accurate as possible, we boost some of the accuracy settings, such as the number of points in $z$ and $k$ used for the power spectrum splines and the number of points at which the kernels are evaluated.

\begin{figure}[h]
    \centering
    \includegraphics[width = 0.95\columnwidth]{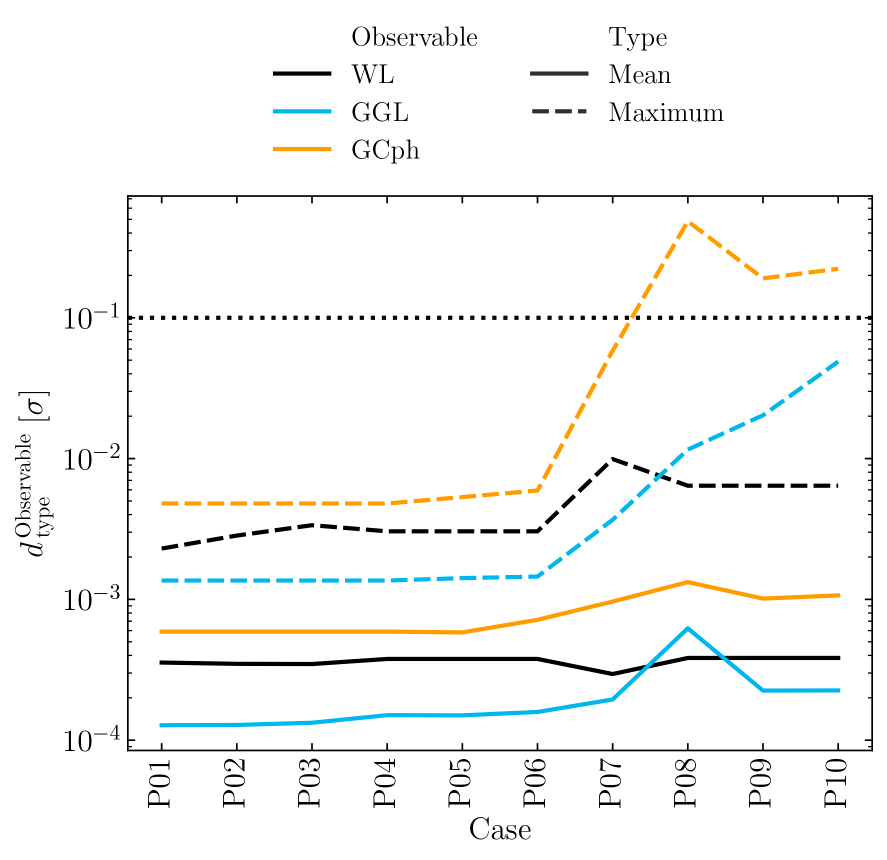}
    \includegraphics[width = 0.95\columnwidth]{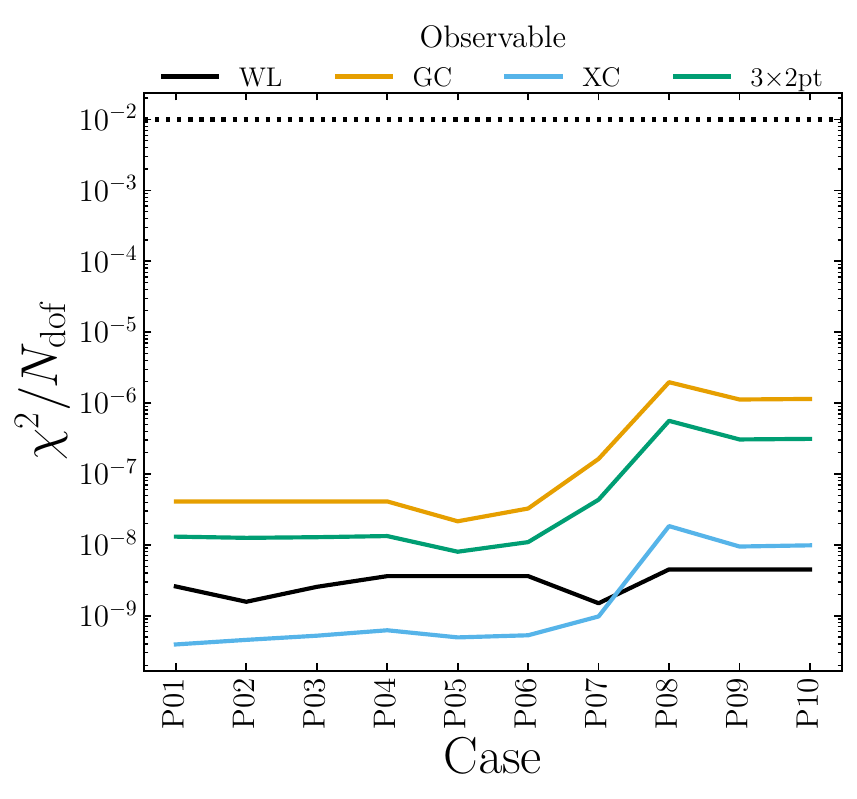}
    \caption{Equivalent of \cref{fig:obs_summary} using \ccl as a benchmark.}
    \label{fig:summary_distance_CCL}
\end{figure}

The results of the comparison, shown in \cref{fig:summary_distance_CCL}, are below the $0.1\,\sigma$ threshold for all the cases investigated, which, as mentioned above, are all the cases without RSD. 

Generally, the distance between \CLOE and \ccl increases with the complexity of the source and lens redshift distribution, as is also the case for the comparison against \life (see \cref{fig:obs_summary}); this is an expected result, and is more prominent for GCph since the relative kernels are much less smooth and have narrower support. The few cases above the threshold are indeed only for the maximum GCph discrepancy, and only concern individual or very small sets of $\ell$ values. As can be seen in the bottom panel of \cref{fig:obs_summary}, such differences do not lead to statistical incompatibilities.

\section{Cosmology with \CLOE+ \class}\label{sec:class_comp}

The results and main conclusions that we draw in this work are obtained using \CLOE interfaced to the Boltzmann solver \camb, through the structure of \cobaya\ 
\citepalias[see][]{Paper2}.
However, \CLOE has been built having in mind the possibility of switching between \camb and \class without any significant change in the pipeline.

At the time of this comparison, the version of \class interfaced does not allow us to use the non-linear recipe we identified as our baseline choice (\texttt{HMCODE2020}), and therefore we cannot apply to the \CLOE+ \class configuration the same validation pipeline we used for \CLOE+ \camb.

Nevertheless, the calculations performed by \CLOE are independent of the choice made on the Boltzmann solver, on which the code relies only to obtain a set of basic cosmological ingredients, used in the calculations. For such a reason, to assess the agreement between \CLOE+ \camb and \CLOE+ \class, it is enough to compare these functions in the two cases, as all subsequent calculations will be identical, whatever the choice for the Boltzmann solver is.

These basic cosmological ingredients necessary for \CLOE are:
\begin{itemize}
    \item the normalised Hubble rate $E(z)\equiv H(z)/H_0$;
    \item the comoving distance $r(z)$;
    \item the linear matter power spectrum $P_{\rm m}^{\rm LIN}(k,z)$.
\end{itemize}

Through these functions, \CLOE computes other derived cosmological quantities, such as the growth factor $g_+(z)$ and the growth rate $f(z)$, and later combines them all to compute the observables 
\citepalias[for details, see][]{EP-CLOE1}. 
While the linear matter power spectrum $P_{\delta\delta}^{\rm LIN}$ is not used directly in any calculation, this is the input for the non-linear methods used in \CLOE, and therefore it constitutes one of the relevant cosmological quantities.

\begin{figure}[h]
    \centering
    \includegraphics[width = 0.95\columnwidth]{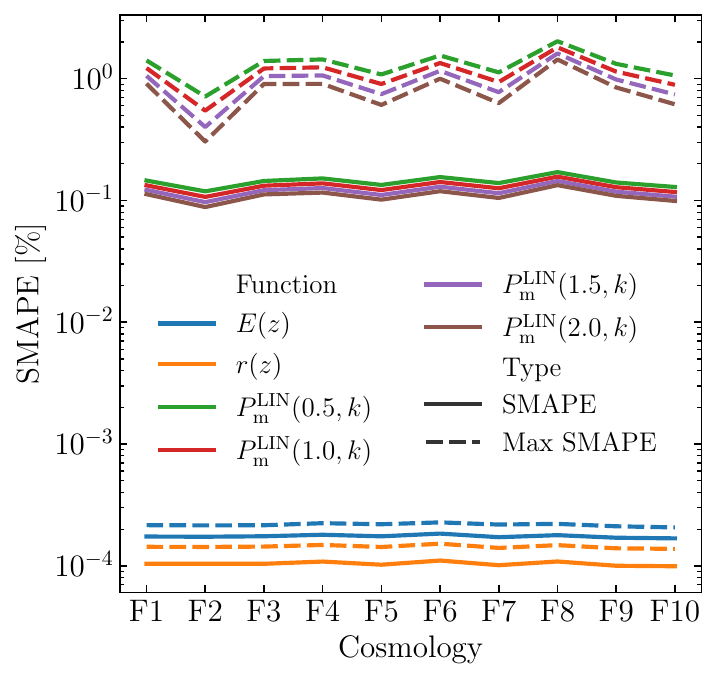}
    \caption{Summary SMAPE values for the comparison between the cosmological function extracted from the Boltzmann solver in the \CLOE+\camb and \CLOE+\class configurations.}
    \label{fig:CAMBvsCLASS}
\end{figure}

In \cref{fig:CAMBvsCLASS}, we show the SMAPE obtained for these four main functions, with the comparison for the last function performed at four different values of the redshift.

We find no significant difference in these functions when switching between the two Boltzmann solvers, with all the differences being at most at the level of one per cent. Such a result is not surprising, given that the structure choices made in developing \CLOE were also aimed at removing any dependence on the choice of the solver. Nevertheless, the results found highlight how, also in the part of the code that directly interfaces with the Boltzmann solvers, the differences are not significant.

\end{appendix}

\label{LastPage}
\end{document}